\documentstyle[eqsecnum,epsfig,aps,amsmath,floats]{revtex} 

\def\met{\mbox{${\hbox{$E$\kern-0.6em\lower-.1ex\hbox{/}}}_T~$}} 
\def\D0{D\O}                            
\def\vmet{\mbox{${\hbox{$\vec{E}$\kern
    -0.6em\lower-.1ex\hbox{/}}}_{T}$}}  

\begin{document}
\pagestyle{myheadings} 
\draft   
\lefthyphenmin=2
\righthyphenmin=3

%
%
\title{
Subjet Multiplicity of Gluon and Quark 
Jets Reconstructed 
with the $k_{\perp}$ Algorithm in
$p\bar{p}$ Collisions
}

%
\author{                                                                      
V.M.~Abazov,$^{23}$                                                           
B.~Abbott,$^{57}$                                                             
A.~Abdesselam,$^{11}$                                                         
M.~Abolins,$^{50}$                                                            
V.~Abramov,$^{26}$                                                            
B.S.~Acharya,$^{17}$                                                          
D.L.~Adams,$^{59}$                                                            
M.~Adams,$^{37}$                                                              
S.N.~Ahmed,$^{21}$                                                            
G.D.~Alexeev,$^{23}$
% ...
\\                                                                            
\vskip 0.30cm                                                                 
\centerline{(D\O\ Collaboration)}                                             
\vskip 0.30cm                                                                 
}                                                                             
\address{                                                                     
\centerline{$^{1}$Universidad de Buenos Aires, Buenos Aires, Argentina}       
\centerline{$^{2}$LAFEX, Centro Brasileiro de Pesquisas F{\'\i}sicas,         
                  Rio de Janeiro, Brazil}                                     
\centerline{$^{3}$Universidade do Estado do Rio de Janeiro,                   
                  Rio de Janeiro, Brazil}                                     
\centerline{$^{4}$Institute of High Energy Physics, Beijing,                  
                  People's Republic of China}                                 
\centerline{$^{5}$Universidad de los Andes, Bogot\'{a}, Colombia}             
\centerline{$^{6}$Charles University, Center for Particle Physics,            
                  Prague, Czech Republic}                                     
\centerline{$^{7}$Institute of Physics, Academy of Sciences, Center           
                  for Particle Physics, Prague, Czech Republic}               
\centerline{$^{8}$Universidad San Francisco de Quito, Quito, Ecuador}         
\centerline{$^{9}$Institut des Sciences Nucl\'eaires, IN2P3-CNRS,             
                  Universite de Grenoble 1, Grenoble, France}                 
\centerline{$^{10}$CPPM, IN2P3-CNRS, Universit\'e de la M\'editerran\'ee,     
                  Marseille, France}                                          
\centerline{$^{11}$Laboratoire de l'Acc\'el\'erateur Lin\'eaire,              
                  IN2P3-CNRS, Orsay, France}                                  
\centerline{$^{12}$LPNHE, Universit\'es Paris VI and VII, IN2P3-CNRS,         
                  Paris, France}                                              
\centerline{$^{13}$DAPNIA/Service de Physique des Particules, CEA, Saclay,    
                  France}                                                     
\centerline{$^{14}$Universit{\"a}t Mainz, Institut f{\"u}r Physik,            
                  Mainz, Germany}                                             
\centerline{$^{15}$Panjab University, Chandigarh, India}                      
\centerline{$^{16}$Delhi University, Delhi, India}                            
\centerline{$^{17}$Tata Institute of Fundamental Research, Mumbai, India}     
\centerline{$^{18}$Seoul National University, Seoul, Korea}                   
\centerline{$^{19}$CINVESTAV, Mexico City, Mexico}                            
\centerline{$^{20}$FOM-Institute NIKHEF and University of                     
                  Amsterdam/NIKHEF, Amsterdam, The Netherlands}               
\centerline{$^{21}$University of Nijmegen/NIKHEF, Nijmegen, The               
                  Netherlands}                                                
\centerline{$^{22}$Institute of Nuclear Physics, Krak\'ow, Poland}            
\centerline{$^{23}$Joint Institute for Nuclear Research, Dubna, Russia}       
\centerline{$^{24}$Institute for Theoretical and Experimental Physics,        
                   Moscow, Russia}                                            
\centerline{$^{25}$Moscow State University, Moscow, Russia}                   
\centerline{$^{26}$Institute for High Energy Physics, Protvino, Russia}       
\centerline{$^{27}$Lancaster University, Lancaster, United Kingdom}           
\centerline{$^{28}$Imperial College, London, United Kingdom}                  
\centerline{$^{29}$University of Arizona, Tucson, Arizona 85721}              
\centerline{$^{30}$Lawrence Berkeley National Laboratory and University of    
                  California, Berkeley, California 94720}                     
\centerline{$^{31}$University of California, Davis, California 95616}         
\centerline{$^{32}$California State University, Fresno, California 93740}     
\centerline{$^{33}$University of California, Irvine, California 92697}        
\centerline{$^{34}$University of California, Riverside, California 92521}     
\centerline{$^{35}$Florida State University, Tallahassee, Florida 32306}      
\centerline{$^{36}$Fermi National Accelerator Laboratory, Batavia,            
                   Illinois 60510}                                            
\centerline{$^{37}$University of Illinois at Chicago, Chicago,                
                   Illinois 60607}                                            
\centerline{$^{38}$Northern Illinois University, DeKalb, Illinois 60115}      
\centerline{$^{39}$Northwestern University, Evanston, Illinois 60208}         
\centerline{$^{40}$Indiana University, Bloomington, Indiana 47405}            
\centerline{$^{41}$University of Notre Dame, Notre Dame, Indiana 46556}       
\centerline{$^{42}$Iowa State University, Ames, Iowa 50011}                   
\centerline{$^{43}$University of Kansas, Lawrence, Kansas 66045}              
\centerline{$^{44}$Kansas State University, Manhattan, Kansas 66506}          
\centerline{$^{45}$Louisiana Tech University, Ruston, Louisiana 71272}        
\centerline{$^{46}$University of Maryland, College Park, Maryland 20742}      
\centerline{$^{47}$Boston University, Boston, Massachusetts 02215}            
\centerline{$^{48}$Northeastern University, Boston, Massachusetts 02115}      
\centerline{$^{49}$University of Michigan, Ann Arbor, Michigan 48109}         
\centerline{$^{50}$Michigan State University, East Lansing, Michigan 48824}   
\centerline{$^{51}$University of Nebraska, Lincoln, Nebraska 68588}           
\centerline{$^{52}$Columbia University, New York, New York 10027}             
\centerline{$^{53}$University of Rochester, Rochester, New York 14627}        
\centerline{$^{54}$State University of New York, Stony Brook,                 
                   New York 11794}                                            
\centerline{$^{55}$Brookhaven National Laboratory, Upton, New York 11973}     
\centerline{$^{56}$Langston University, Langston, Oklahoma 73050}             
\centerline{$^{57}$University of Oklahoma, Norman, Oklahoma 73019}            
\centerline{$^{58}$Brown University, Providence, Rhode Island 02912}          
\centerline{$^{59}$University of Texas, Arlington, Texas 76019}               
\centerline{$^{60}$Texas A\&M University, College Station, Texas 77843}       
\centerline{$^{61}$Rice University, Houston, Texas 77005}                     
\centerline{$^{62}$University of Virginia, Charlottesville, Virginia 22901}   
\centerline{$^{63}$University of Washington, Seattle, Washington 98195}       
}                                                                             


\maketitle

%
%
\begin{abstract}
The D\O\ Collaboration has studied
for the first time
the properties of hadron-collider jets
reconstructed with a successive-combination 
algorithm based on relative
transverse momenta ($k_{\perp}$)
of energy clusters.
Using the standard value $D = 1.0$ of the jet-separation
parameter in the $k_{\perp}$ algorithm,
we find that the
$p_T$ of such jets is higher than the
$E_T$ of matched jets reconstructed with cones of radius ${\cal R} = 0.7$,
by about $5$ (8) GeV at $p_T \approx 90$ (240) GeV.
To examine internal jet structure, 
the $k_{\perp}$ algorithm is applied within $D=0.5$ jets 
to resolve any subjets.
The multiplicity of subjets in jet
samples at $\sqrt{s} = 1800$ GeV and 630 GeV is
extracted separately for gluons ($M_g$)
and quarks ($M_q$), and
the ratio of average
subjet multiplicities 
in gluon and quark jets is 
measured as
$
\frac{\langle M_{g} \rangle -1} {\langle M_{q} \rangle -1}
=1.84\pm
0.15\;(\text{stat.})\pm ^{0.22}_{0.18} \;(\text{sys.})
$.
This ratio is in agreement with the
expectations from
the {\sc HERWIG} Monte Carlo
event generator and a resummation calculation,
and with observations in $e^+e^-$ annihilations,
and is close to the
naive prediction for the ratio 
of color charges of $C_A/C_F = 9/4 = 2.25$.
\end{abstract}

\begin{center}
(submitted to Phys. Rev. D)
\end{center}


\vfill\eject

\twocolumn

\section{Introduction}
The production of gluons and quarks in 
high-energy
collisions,
and their development into the jets of particles observed in experiments,
is usually described by the theory of Quantum Chromodynamics (QCD).
In perturbative QCD, a produced parton (gluon or quark) emits gluon radiation,
with each subsequent emission carrying off a fraction of
the original parton's energy and momentum. 
The probability for a gluon to radiate
a gluon is proportional to the color factor $C_A = 3$, while gluon radiation
from a quark is proportional to the color factor $C_F = 4 / 3$.
In the asymptotic limit, in which the radiated gluons carry a small fraction
of the original parton's momentum, and neglecting
the splitting of gluons to quark-antiquark pairs
(whose probability is proportional to the color factor $T_R$ = 1/2),
the average number of objects radiated by a gluon
is expected to be a factor $C_A / C_F = 9 / 4$
higher than the number of objects radiated by a quark~\cite{Ellis}.
In general, it is expected that a gluon
will yield more 
particles with a softer momentum distribution,
relative to a quark~\cite{lep,leprev}.

Although gluon jets are expected to 
dominate the final state of proton-antiproton ($p\bar{p}$)
collisions at high energies,
quark jets make up a significant fraction of the jet cross section
at high $x_T = \frac {2 p_T} {\sqrt{s}}$, 
where $\sqrt{s}$ is the total energy of the $p\bar{p}$ system,
and $p_T$ is the jet momentum transverse to the hadron-beam 
direction.
The ability to distinguish gluon jets from quark jets
would provide a powerful tool in the study of hadron-collider physics.
To date, however, 
there has been only little experimental verification
that gluon jets 
produced in hadron collisions
display characteristics
different from quark jets~\cite{ua2,ua1,alljets,snihur,safonov}.
For fixed $p_T$, we analyze
the internal structure of 
jets at  $\sqrt{s} = 1800$ GeV and 630 GeV
by 
resolving jets within
jets 
({\em subjets})~\cite{snihur,cat_sub1,opal,sey_sub1,astur,aleph,sey_sub0,amy,Abreu:1998ve,sey_sub2}.
Using the expected fractions of gluon and quark jets
at each  $\sqrt{s}$, we measure the multiplicity 
of 
subjets
in gluon and in quark jets.
The results are presented as a ratio of average multiplicities
$r = \frac {\langle M_g \rangle -1} {\langle M_q \rangle -1}$
of subjets in gluon jets to quark jets.
This measured ratio is compared to 
that observed in $e^+e^-$ annihilations~\cite{aleph,Abreu:1998ve},
to predictions 
of a resummed calculation~\cite{sey_sub1,sey_sub0,sey_sub2}, and 
to the {\sc HERWIG}~\cite{hw} Monte Carlo generator of jet events.

The D\O\ detector~\cite{d0},
described briefly in Sec.~\ref{subs:det},
is well-suited to studying properties of jets.
A jet algorithm associates the large number of particles 
produced in a hard-scattering process
with the quarks and gluons of QCD.
We define jets with a successive-combination 
algorithm~\cite{cat93,cat92,ES,framedpf} based on relative
transverse momenta ($k_{\perp}$) of energy clusters, 
described in Sec.~\ref{subs:alg}.
In this paper, we present the first 
measurement of jet properties
using the $k_{\perp}$ (sometimes written $k_T$)
algorithm at a hadron collider.
The momentum calibration of jets in the $k_{\perp}$ algorithm 
is outlined in Sec.~\ref{subs:jes}, followed 
in Sec.~\ref{subs:comp} by a simple comparison
with jets defined with the fixed-cone algorithm.
To study jet structure, the $k_{\perp}$ algorithm 
is then applied within the jet
to resolve subjets, as described in Sec.~\ref{subs:sub}.
In $e^+e^-$ annihilations, the number of subjets in gluon jets was shown
to be larger than in quark jets~\cite{aleph,Abreu:1998ve}.
In $p\bar{p}$ collisions, identifying gluon and quark jets
is more complicated than in $e^+e^-$ annihilations.
We approach this issue by
comparing central jet samples at $\sqrt{s} = 1800$ GeV and 630 GeV, 
with the samples described in
Sec.~\ref{subs:sel}.  For moderate jet $p_T$ (55 -- 100 GeV),
the $\sqrt{s} = 1800$ GeV sample is gluon-enriched,
and the $\sqrt{s} =  630$ GeV sample is quark-enriched.
Section \ref{subs:ext} describes a simple method developed 
to extract the separate subjet multiplicity for gluon and for quark jets. 
The method does not tag individual jets, but
instead, we
perform a statistical analysis of the samples at 
$\sqrt{s} = 1800$ GeV and
$630$ GeV~\cite{Acton:1993jm}.
The method requires the relative mix of quarks and gluons
in the two data samples, which is derived from a Monte Carlo
event generator that uses the parton
distribution functions~\cite{cteq,mrs}, measured primarily in deep
inelastic scattering. 
Subsequent sections describe the measurement of the subjet multiplicity in
D\O\ data 
and Monte Carlo simulations, the corrections used in the procedure, and the 
sources of systematic uncertainty.
We conclude with comparisons to previous experimental and theoretical
studies.

\section{D\O\ detector}
\label{subs:det}
\begin{figure} 
\epsfxsize=2.875in
\hspace {-0.3cm}
\epsfbox{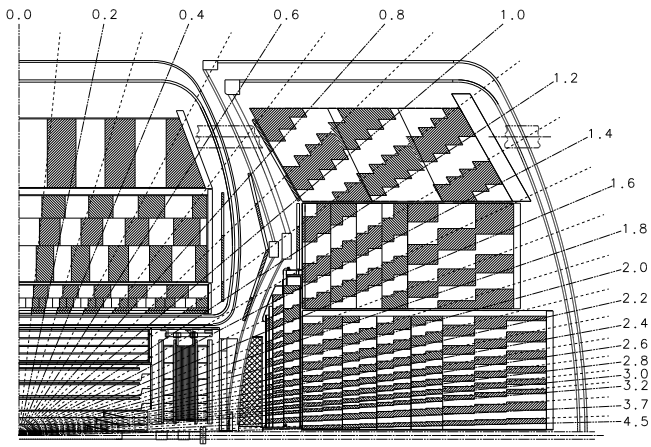}
\vspace{0.5cm}
\caption{
One quadrant of the D\O\ calorimeter and drift chambers,
projected in the $r - z$ plane.
Radial lines illustrate the detector pseudorapidity
and the pseudo-projective geometry of the calorimeter towers. 
Each tower has size
$\Delta \eta \times \Delta \phi = 0.1 \times 0.1$.
}
\label{fig:cal_side_eta}
\end{figure}

D\O\ is a multipurpose detector designed to study $p\bar{p}$
collisions at the Fermilab Tevatron Collider.
A full description of the D\O\ detector can be found in Ref.~\cite{d0}.
The primary detector components for jet measurements at D\O\
are the excellent compensating calorimeters. 
The D\O\ calorimeters
use liquid-argon as the active medium to 
sample the ionization energy produced in electromagnetic and hadronic showers.
The elements of the calorimeter systems are housed in
three cryostats.
The central calorimeter (CC) covers the 
region $|\eta| < 1.0$, while the symmetric end calorimeters (EC)
extend coverage to $|\eta| < 4.2$,
where the pseudorapidity $\eta = - \ln \tan \theta / 2$ is defined 
in terms of the polar angle $\theta$ with respect to the proton-beam direction $z$.
Each system is divided into an electromagnetic (EM), fine hadronic (FH),
and coarse hadronic (CH) sections.
The EM and FH use uranium absorber plates as the passive medium, 
and the CH uses either copper (CC) or stainless
steel (EC).
Copper readout pads are 
centered in the liquid-argon gaps between 
the absorber plates.
Radially, the electromagnetic sections
are 21 radiation lengths deep, divided into
4 readout layers.
The hadronic calorimeters are 7--11 nuclear interaction lengths
deep, with up to 4 layers.
The entire calorimeter is segmented into towers, of typical size 
$\Delta \eta \times \Delta \phi = 0.1 \times 0.1$,
projected towards the nominal 
$p\bar{p}$ interaction point in the center of the detector,
where $\phi$ is the azimuthal angle 
about the $z$ axis.
Figure~\ref{fig:cal_side_eta} shows a schematic view
of one quadrant of the D\O\ calorimeter in the $r-z$ plane,
where $r$ is the distance from the origin in the plane transverse
to the beam axis.
Each layer in a calorimeter tower is called a cell,
and yields an individual energy sampling.
Energy deposited in the calorimeters by particles
from $p\bar{p}$
collisions are used to reconstruct jets.
The transverse energy resolution of jets for data at $\sqrt{s} = 1800$ GeV 
can be parameterized as~\cite{iain}:
\begin{equation}
(\sigma(E_T) / E_T)^2 \approx  6.9 / E_T^2 + 0.5 / E_T + 0.001,
\end{equation}
with $E_T$ in GeV.

\begin{figure} 
\epsfxsize=3.375in
\epsfbox{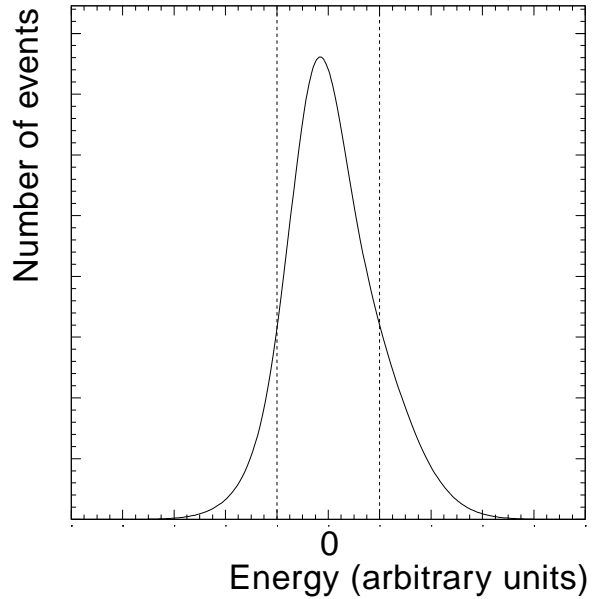}
\vspace{0.2cm}
\caption{Illustration of the pedestal energy distribution in a 
calorimeter cell (solid line), stemming from uranium noise.
The mean value is defined to be zero, and the peak occurs
at negative values.  Removal of the portion between the
vertical dashed lines (a symmetric window about the mean)
yields a positive mean for the remaining distribution.
}
\label{fig:u_decay}
\end{figure}

In the analysis of jet structure,
we are interested in the distribution of energy within jets.
Apart from the energy of particles produced
in a hard-scattering event,
the cells of the D\O\ calorimeter are sensitive
to three additional sources of energy that
contribute to a jet.
The first, called uranium noise, 
is a property of the detector material.
The decay of radioactive uranium nuclei in the calorimeter
can produce energy in a given cell,
even in the absence of a particle flux.
For each cell, a distribution of this 
pedestal energy is measured in a series of calibration
runs without beams in the accelerator.
The pedestal 
distribution due to uranium noise is asymmetric, with a longer high-end tail,
as illustrated in Fig.~\ref{fig:u_decay}.
During normal data-taking, 
the mean pedestal energy is subtracted online
from the energy measured in a hard-scattering event.
To save processing time and reduce the event size, 
a zero-suppression circuit is used,
whereby
cells containing energy within a symmetric window about 
the mean pedestal count are not read out.
Since the pedestal distribution of each cell is asymmetric, zero-suppression 
causes upward fluctuations
in measured cell energies more often than downward fluctuations.
In the measurement of a hard-scattering event,
the net impact is an increased multiplicity of readout cells and
a positive offset to their initial energies.

\begin{figure} 
\epsfxsize=3.375in
\hspace {-0.2cm}
\epsfbox{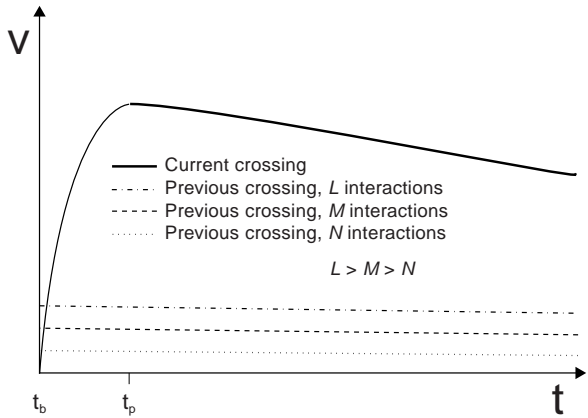}
\vspace{0.2cm}
\caption{Schematic of signal voltage in 
a calorimeter cell as a function of time.
The solid line represents the contribution for a given event 
(the ``current'' $p\bar{p}$ bunch crossing).
In the absence of previous bunch crossings,
the cell is sampled correctly at $t_b$, just before a crossing,
to establish a base voltage, and at $t_p$,
to establish a peak voltage.
The voltage difference
$\Delta V = V(t_p) - V(t_b)$ is proportional to
the initial energy deposited in the cell.
The dashed lines show example
contributions from a previous bunch crossing
containing three different numbers of $p\bar{p}$ interactions.
The observed signal is the sum of the signals from the current
and previous crossings.
(The figure is not to scale.)
}
\label{fig:cal_signal2}
\end{figure}

There are two other 
environmental
effects that contribute to
the energy offset of calorimeter cells.
The first is extra energy from multiple $p\bar{p}$ interactions
in the same accelerator-bunch crossing,
and this depends on the instantaneous luminosity.
To clarify the second effect, called pile-up,
we turn to how calorimeter cells are sampled,
as is illustrated in Fig.~\ref{fig:cal_signal2}.
The maximum drift time for ionization electrons produced in the liquid-argon 
to reach the copper readout pad of a calorimeter cell is about 450 ns.
The collected electrons produce an electronic signal
that
is sampled at the time of the bunch crossing (base),
and again 2.2 $\mu$s later (peak).
The difference in voltage between the two samples (peak relative to base) 
defines
the initial energy count in a given cell.  
Because the signal fall-time ($\sim \! 30$ $\mu$s) is longer than
the accelerator bunch spacing (3.5 $\mu$s),
the base and peak voltages are measured with respect to
a reference level that
depends
on previous bunch crossings.  
The signal from the current bunch crossing is therefore
piled on top of the decaying signal from previous crossings.
When a previous bunch crossing leaves energy in
a particular cell, that cell's 
energy count will therefore be reduced on average,
after the baseline subtraction.

\section{$\lowercase{k}_{\perp}$ jet algorithm}
\label{subs:alg}
Jet algorithms assign {\em particles} produced in
high-energy collisions to jets.
The {\em particles} correspond to observed energy depositions in a calorimeter,
or to final state particles generated in a Monte Carlo event.
Typically, such objects 
are first organized into preclusters (defined below),
before being processed through the jet algorithms:
The jet algorithms therefore do not depend on the nature of the particles.
We discuss two jet algorithms in this paper: the $k_{\perp}$ and cone 
jet algorithms, with emphasis on the former.

In the $k_{\perp}$ jet algorithm,
pairs of particles are merged successively 
into jets,
in an order corresponding to increasing relative transverse momentum.
The algorithm contains a single parameter $D$
(often called $R$ in some references),
which controls the cessation of merging.
Every particle in the event is assigned to a single $k_{\perp}$ jet.

In contrast, the fixed-cone algorithm~\cite{snowmass} associates into a jet
all particles with trajectories within an
area $A = \pi {\cal R}^2$, 
where the parameter ${\cal R}$
is the radius of a cone in $(\eta,\phi)$ space.
The D\O\ fixed-cone algorithm~\cite{iain,run2qcd} is an iterative algorithm,
starting with cones centered on the most energetic
particles in the event (called seeds).
The energy-weighted centroid of a cone is defined by:
\begin{equation}
\eta^C = \frac {\sum_{i} E_T^i \eta^i} {\sum_{i} E_T^i},
\phi^C = \frac {\sum_{i} E_T^i \phi^i} {\sum_{i} E_T^i},
\end{equation}
where the sum is over all particles $i$ in the cone.
The centroids are used iteratively as centers for new cones in 
$(\eta,\phi)$ space.
A jet axis is defined when a cone's centroid and geometric center coincide.
The fixed-cone jet algorithm
allows cones to overlap,
and any single particle can belong to two or more jets.
A second parameter, and additional steps, are 
needed to determine if
overlapping cones should be split or merged~\cite{rsep}.

The $k_{\perp}$ jet algorithm offers several advantages over the 
fixed-cone jet algorithms, which are widely used at hadron colliders.
Theoretically, the $k_{\perp}$ algorithm is infrared-safe and 
collinear-safe to 
all orders of calculation~\cite{cat93,run2qcd}.
The same algorithm can be applied to partons generated 
from fixed-order or resummation 
calculations in QCD, particles in a Monte Carlo event generator,
or tracks or energy depositions in a detector.

The  $k_{\perp}$ jet
algorithm is specified in Sec.~\ref{subs:clu}.
In Sec.~\ref{subs:pre}, we describe the preclustering algorithm,
the goal of which is to reduce
the detector-dependent aspects of jet clustering
(e.g., energy thresholds or calorimeter segmentation).
The momentum calibration of $k_{\perp}$ jets
is presented in Sec.~\ref{subs:jes}.
In Sec.~\ref{subs:comp},
jets reconstructed using the $k_{\perp}$ algorithm
are compared
to jets reconstructed with
the fixed-cone algorithm.
In Sec.~\ref{subs:sub}, we indicate how subjets are defined in the
$k_{\perp}$ algorithm.

\subsection{Jet clustering}
\label{subs:clu}
There are several variants of the $k_{\perp}$ jet-clustering algorithm 
for hadron colliders~\cite{cat93,cat92,ES}.
The main differences concern how particles are merged together
and when the clustering stops.
The different types of merging, or recombination, 
schemes were investigated in Ref.~\cite{cat93}.
D\O\ chooses the scheme that corresponds
to four-vector addition of momenta, because~\cite{run2qcd}:
\begin{enumerate}
\item
it is conceptually simple;
\item
it corresponds to the scheme used in the $k_{\perp}$ algorithm in $e^+e^-$
annihilations~\cite{aleph,Abreu:1998ve}; 
\item
it has no energy defect~\cite{kosower}, 
a measure of perturbative stability
in the analysis of transverse energy density within jets; and 
\item
it is better suited~\cite{frame} to the missing transverse energy
calculation in the jet-momentum
calibration method used by D\O.
\end{enumerate}
To stop clustering, D\O\ has adopted the proposal~\cite{ES}
that halts clustering when all the jets are separated by
$\Delta{\cal R} > D$.  This rule is simple,
and maintains a similarity with cone algorithms for hadronic collisions.
The value $D = 1.0$ treats initial-state radiation in the 
same way as final-state radiation~\cite{sey_sub1,sey_pbarp}.

The jet algorithm starts with a list of preclusters
as defined in the next section.
Initially, each precluster is assigned a momentum four-vector
$(E,{\bf p}) = E_{{\rm precluster}}(1, \sin \theta \cos \phi , \sin \theta \sin \phi , \cos \theta)$,
written in terms of the precluster angles $\theta$ and $\phi$.
The execution of the jet algorithm involves:

1. Defining for each object $i$ in the list:
\[
d_{ii} \equiv p_{T,i}^2 = p_{x,i}^2 + p_{y,i}^2,
\]
and for each pair $(i,j)$ of objects:
\begin{eqnarray}
d_{ij} & \equiv & \min\left[ p_{T,i}^2,p_{T,j}^2 \right] \frac{ \Delta{\cal R}_{ij}^2} {D^2} \nonumber \\
& = & \min\left[ p_{T,i}^2,p_{T,j}^2 \right] \frac{(\eta_i - \eta_j)^2 + (\phi_i - \phi_j)^2} {D^2},
\label{eq:dij}
\end{eqnarray}
where $D$ is the stopping parameter of the jet algorithm.
For $D = 1.0$ and $\Delta{\cal R}_{ij} \ll 1$, $d_{ij}$ reduces to the square
of the relative transverse momentum
($k_{\perp}$) between objects.

\begin{figure}
\epsfxsize=3.375in
\epsfbox{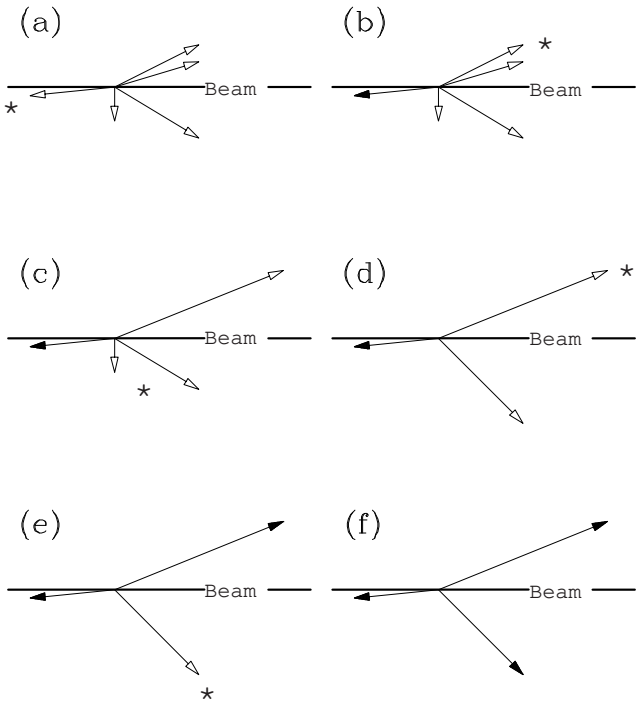}
\vspace{0.2cm}
\caption{A simplified example of the final state of a collision 
between two hadrons.  
(a) The particles in the event (represented by arrows)
comprise a list of objects.
(b-f) Solid arrows represent
the final jets reconstructed by the $k_{\perp}$ algorithm,
and open arrows represent objects not yet assigned to jets.
The five diagrams show successive iterations of the algorithm.
In each diagram, a jet is either defined (when it is
well-separated from all other objects), or two objects are merged
(when they have small relative $k_{\perp}$).
The asterisk labels the relevant object(s) at each step.
}
\label{fig:kt_example}
\end{figure}

2. If the minimum of all possible $d_{ii}$ and $d_{ij}$ is a $d_{ij}$,
then replacing objects $i$ and $j$ by their merged object $(E_{ij},{\bf p}_{ij})$, where
\[
E_{ij} = E_i + E_j
\]
\[
{\bf p}_{ij} = {\bf p}_i + {\bf p}_j.
\]
And if the minimum is a $d_{ii}$, then 
removing object $i$ from the list and defining it to be a jet.

3. Repeating Steps 1 and 2 when there are any objects left in the list.

The algorithm produces a list of jets, each separated by $\Delta{\cal R} > D$.
Figure~\ref{fig:kt_example} illustrates how the $k_{\perp}$ 
algorithm successively merges
the particles in a simplified diagram of a hadron collision.

\subsection{Preclustering}
\label{subs:pre}
In the computer implementation of the $k_{\perp}$ jet 
algorithm, the processing time is proportional to $N^3$,
where $N$ is the number of particles (or energy signals)
in the event~\cite{cat93}.
The zero-suppression circuit 
reduces the number of calorimeter cells that have to be read out
in each event.  To reduce this further,
we employ a preclustering algorithm.  The procedure
assigns calorimeter cells
(or particles in a Monte Carlo event generator)
to preclusters, suitable for input to the jet-clustering
algorithm.
In essence, calorimeter cells are collapsed into towers,
and towers are merged if they are close together in
$(\eta,\phi)$ space or if they have small 
$p_T$.
Monte Carlo studies have shown that such preclustering reduces the
impact of ambiguities due to
calorimeter showering and finite segmentation,
especially on the reconstructed internal jet substructure.
For example, when a single particle strikes the boundary between two 
calorimeter towers, it can produce two clusters of energy.
Conversely, two collinear particles
will often shower in a single calorimeter tower.
In both cases, there is a potential discrepancy
in the number of energy clusters found at
the calorimeter level and the particle level.
Preclustering at both the calorimeter and at the particle level
within a radius 
larger than the calorimeter segmentation
integrates over such discrepancies.

\begin{figure} 
\epsfxsize=3.375in
\epsfbox{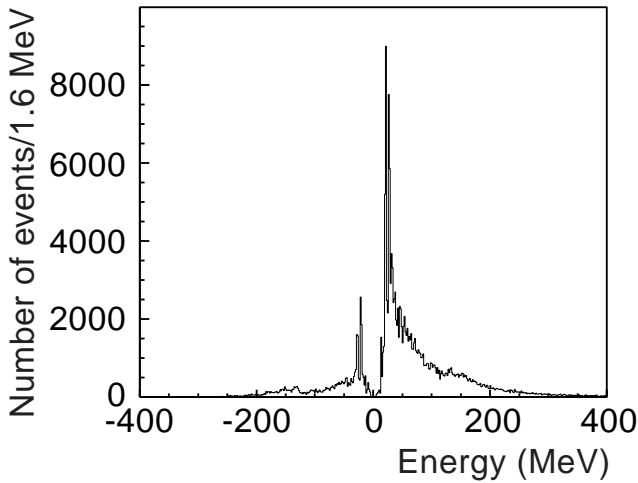}
\vspace{0.2cm}
\caption{
Mean energies in calorimeter cells 
for a sample of minimum-bias events.  
The contribution from instrumental effects is included,
which occasionally leads to negative energy readings.
For each cell, the energy distribution illustrated in Fig.~\ref{fig:u_decay}
is fitted to a Gaussian.
Before readout, the zero-suppression circuit in each cell's electronics
sets to zero energy the channels in
a symmetric window about the mean pedestal.
These channels are not read out,
causing the dip observed near zero.
}
\label{fig:ktcl_page3_fig1}
\end{figure}

The preclustering algorithm consists of the following six steps:

1. Starting from a list of populated calorimeter cells in an event,
remove any cells with $E_T < -0.5$ GeV.
Cells with such
negative $E_T$ --- rarely observed in 
minimum-bias\footnote{The minimum-bias trigger requires a 
coincidence signal
in the scintillating-tile hodoscopes~\cite{d0} located near the beampipe.} 
events (see Fig.~\ref{fig:ktcl_page3_fig1}) ---
are considered spurious.

2. For each calorimeter cell centered at some $(\theta,\phi)$ 
relative to the primary interaction vertex,
define its pseudorapidity:
\[
\eta = - \ln \tan \frac{\theta}{2}.
\]

3. For each calorimeter tower $t$, sum the transverse 
energy of cells $c$ in that tower:
\[
E_T^{t} = \sum_{c \: \epsilon \: t} E_c \sin \theta_{c},
\]
where $E_c$ is the energy deposited in cell $c$.

4. Starting at the extreme negative value of $\eta$ and $\phi = 0$, 
combine any neighboring towers into preclusters such that no
two preclusters are within 
$\Delta{\cal R}^{{\rm pre}} = \sqrt{ \Delta \eta^2 + \Delta \phi^2} = 0.2$.
The combination follows the Snowmass prescription~\cite{snowmass}:
\[
E_T = E_{T,i} + E_{T,j}
\]
\[
\eta = \frac{E_{T,i}\eta_i + E_{T,j}\eta_j} {E_{T,i} + E_{T,j}}
\]
\[
\phi = \frac{E_{T,i}\phi_i + E_{T,j}\phi_j} {E_{T,i} + E_{T,j}}.
\]
The procedure evolves in the direction of increasing $\phi$,
and then increasing $\eta$.

5. Because of
pile-up in the calorimeter,
precluster energies can fluctuate
in both positive and negative directions.
Preclusters that have negative transverse 
energy $E_T = E_{T-} < 0$, are redistributed to $k$ neighboring
preclusters in the following way.  
Given a negative $E_T$ precluster with 
$(E_{T-},\eta_{-},\phi_{-})$,
we define a square $S$ of size 
$(\eta_{-} \pm 0.1) \times (\phi_{-} \pm 0.1)$.
When the following holds:
\begin{equation}
\sum_{k \: \epsilon \: S} E_{T,k}(\eta,\phi) > |E_{T-}|,
\label{eq:pre}
\end{equation}
where only preclusters with positive $E_T$
that are located within 
the square $S$ are included in the sum,
then $E_{T-}$ is 
redistributed to the positive preclusters in the square,
with each such precluster $k$ absorbing a fraction 
\[
\frac {E_{T,k}} {{\displaystyle \sum_{k \: \epsilon \: S}} E_{T,k}}
\]
of the negative $E_T$.
If Eq.~(\ref{eq:pre}) is not satisfied, the ``search square'' is increased 
in steps of
$\Delta \eta = \pm 0.1$ and $\Delta \phi = \pm 0.1$,
and another redistribution is attempted.
If redistribution still fails for a square of
$(\eta_{-} \pm 0.7) \times (\phi_{-} \pm 0.7)$, 
the negative energy precluster is isolated in the calorimeter
and ignored (by setting $E_{T-} = 0$).

\begin{figure}
\epsfxsize=3.375in
\epsfbox{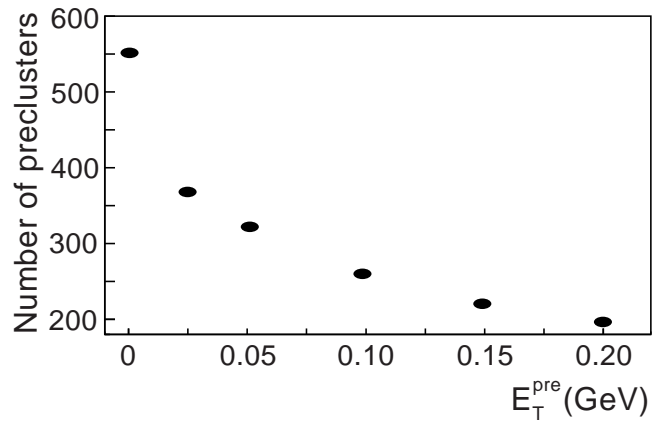}
\vspace{0.2cm}
\caption{The mean number of preclusters per event, 
as a function of the setting of minimum transverse energy
required for preclusters
($E_T^{{\rm pre}}$).
}
\label{fig:ktcl_page6}
\end{figure}

6. Preclusters with $0 < E_T < E_T^{{\rm pre}} = 0.2$ GeV, 
are redistributed to neighboring
preclusters, as specified in Step 5.  To reduce the
overall number of preclusters, we 
also require that the
search square have at least three positive $E_T$ preclusters.
The threshold $E_T^{{\rm pre}}$ was tuned to produce about $200$ preclusters/event
(see Fig.~\ref{fig:ktcl_page6}), in order to fit our constraints for 
processing time.

\subsection{Calibration of jet momentum}
\label{subs:jes}
A correct calibration of jet momentum
reduces overall experimental uncertainties
on jet production.
The calibration at D\O\ also 
accounts for the contribution
of the underlying event 
(momentum transferred as a result of the soft interactions between
the remnant partons of the proton and antiproton).
All such corrections
enter in the relation
between the momentum of a jet measured in the calorimeter $p^{{\rm meas}}$
and the ``true'' jet momentum $p^{{\rm true}}$ ~\cite{jes}
\begin{equation}
p_{{\rm jet}}^{{\rm true}} = 
\frac {p_{{\rm jet}}^{{\rm meas}} - p_O(\eta^{{\rm jet}},{\cal L},p_T^{{\rm jet}})}
{R_{{\rm jet}}(\eta^{{\rm jet}},p^{{\rm jet}})}
\label{eq:jes}
\end{equation}
where $p_O$ denotes an offset correction, 
$R_{{\rm jet}}$ is a correction for the 
response of the calorimeter to jets,
and ${\cal L}$ is the instantaneous luminosity.
A true jet is defined as being composed of only
the final-state particle momenta
from the hard parton-parton scatter
(i.e., before interaction in the calorimeter).
Although 
Eq.~(\ref{eq:jes})
is valid for any jet algorithm,
$p_O$ and the components of $R_{{\rm jet}}$ depend 
on the details of the jet algorithm.
Our calibration procedure attempts to
correct calorimeter-level jets (after interactions in the calorimeter)
to their particle-level 
(before the individual particles interact in the calorimeter),
using
the described  $k_{\perp}$ jet algorithm, 
with $D = 1.0$.
The procedure follows closely that of 
calibration of the fixed-cone jet algorithm~\cite{jes}.
The fixed-cone jet algorithm requires an additional
scale factor in Eq.~(\ref{eq:jes}), but we find no need for
that kind of calorimeter-showering correction in the $k_{\perp}$ 
jet momentum calibration~\cite{frame}.

The offset $p_O$ corresponds to the 
contribution to the
momentum of a reconstructed jet that is not associated
with the hard interaction.
It contains two parts: 
\[
p_O = O_{{\rm ue}} + O_{{\rm zb}},
\]
where $O_{{\rm ue}}$ is the offset due to the underlying event,
and $O_{{\rm zb}}$ is an offset due to the overall detector environment.
$O_{{\rm zb}}$ is attributed to any additional energy in the calorimeter
cells of a jet from the combined effects of uranium noise, multiple interactions, and pile-up.
The contributions of $O_{{\rm ue}}$ and $O_{{\rm zb}}$ to $k_{\perp}$ jets 
are measured separately,
but using similar methods.  The method 
overlays
D\O\ data and Monte Carlo events, as described in what follows.

\begin{figure}
\epsfxsize=3.175in
\epsfbox{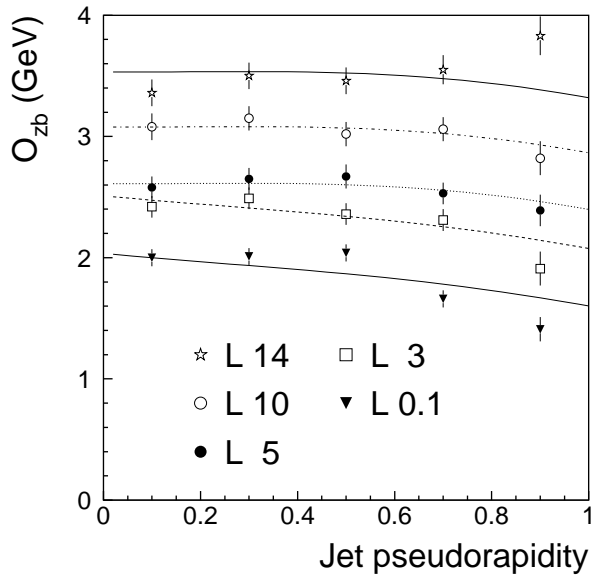}
\vspace{0.2cm}
\caption{
The offset correction $O_{{\rm zb}}$ 
as a function of pseudorapidity of $k_{\perp}$ jet ($D = 1.0$).
The offset $O_{{\rm zb}}$ accounts for the combined effects of pile-up, 
uranium noise, and multiple interactions.
The different sets of points are for events 
with different instantaneous luminosity
${\cal L} \approx 14, 10, 5, 3, 0.1 \times 10^{30} \text{cm}^{-2}\text{s}^{-1}$. 
The curves are fits to the points
at different ${\cal L}$, 
using the same functional form as
employed for the cone algorithm in Ref.~\protect\cite{jes}.
}
\label{fig:ozb_eta}
\end{figure}

The Monte Carlo events are generated by {\sc HERWIG} (version 5.9)~\cite{hw} 
with $2 \rightarrow 2$ parton $p_T$-thresholds 
of 30, 50, 75, 100, and 150 GeV,
and the underlying-event contribution switched off.
The Monte Carlo events are propagated through 
a {\small GEANT}-based~\cite{d0geant} simulation of the D\O\ detector,
which provides a cell-level simulation of the calorimeter response and resolution.
These Monte Carlo events are then passed through the calorimeter-reconstruction
and jet-finding packages, defining the initial sample of jets.
Detector simulation does not include the effects of uranium noise 
nor of the accelerator conditions causing multiple interactions and pile-up.
The total contribution from these three effects is modeled
using zero-bias events, which 
correspond to observations at  random $p\bar{p}$ bunch crossings.
Zero-bias events were recorded by the D\O\ detector at different instantaneous
luminosities in special data-taking runs without the zero-suppression
discussed in Sec.~\ref{subs:det}.
The cell energies in zero-bias events are added cell-by-cell to 
the energies in simulated Monte Carlo jet events.  The summed cell energies are
then zero-suppressed offline, using the pedestals appropriate to
the zero-bias running conditions.  Finally, the summed cell energies are passed through
the calorimeter-reconstruction and jet-finding packages, 
producing a second sample of jets.  The two samples are compared on an
event-by-event basis, associating the jets in events of the two samples
that have their axes separated by $\Delta{\cal R} < 0.5$~\cite{frame}.
The difference in the measured $p_T$ of the corresponding matched jets
is $O_{{\rm zb}}$, and shown in Fig.~\ref{fig:ozb_eta}
as a function of $\eta^{{\rm jet}}$, for different instantaneous luminosities.

The event-overlay method was checked with the fixed-cone jet 
algorithm for ${\cal R}  = 0.7$.  
For jets with 30 GeV $< E_T <$ 50 GeV,
this method gives only  
$14\%$ ($28\%$) smaller offsets 
$[\Delta O_{{\rm zb}} =$ 0.25 (0.39) GeV per jet$]$, at
${\cal L} \approx 5 \: (0.1) \times 10^{30} \text{cm}^{-2}\text{s}^{-1}$
relative to Ref.~\cite{jes}.
Independent of jet $E_T$,
the method used in Ref.~\cite{jes} measures the
$E_T$ per unit $\Delta \eta \times \Delta \phi$
in zero-bias events, and scales the value by the area of the jet cone.
In the event-overlay method,
$O_{{\rm zb}}$ decreases by as much as $40\%$
when the cone-jet transverse energy increases to 125 GeV $< E_T <$  170 GeV.
Approximately $30\%$ of this decrease
can be explained by the $E_T^{{\rm jet}}$-dependence
of the occupancy of cells within cone jets (the fraction of cells 
with significant energy deposition
inside the cone).
The remaining $70\%$ of the 
$O_{{\rm zb}}$ dependence on jet $E_T$ is
assigned as a systematic uncertainty on our method.
Since the observed dependence is less pronounced
in the $k_{\perp}$ jet algorithm,
this error amounts at most
to $15\%$ in the highest jet $p_T$ bin. 
In addition, we include a systematic uncertainty
of 0.2 GeV arising from the fits in 
Fig.~\ref{fig:ozb_eta}.
Using our overlay method for both algorithms,
the offsets $O_{{\rm zb}}$ in the $k_{\perp}$ jet algorithm (with $D = 1.0$)
are generally $50-75\%$ (or about 1 GeV per jet)
larger than in the fixed-cone jet algorithm (with ${\cal R}  = 0.7$)~\cite{frame}.

\begin{figure}
\epsfxsize=3.375in
\epsfbox{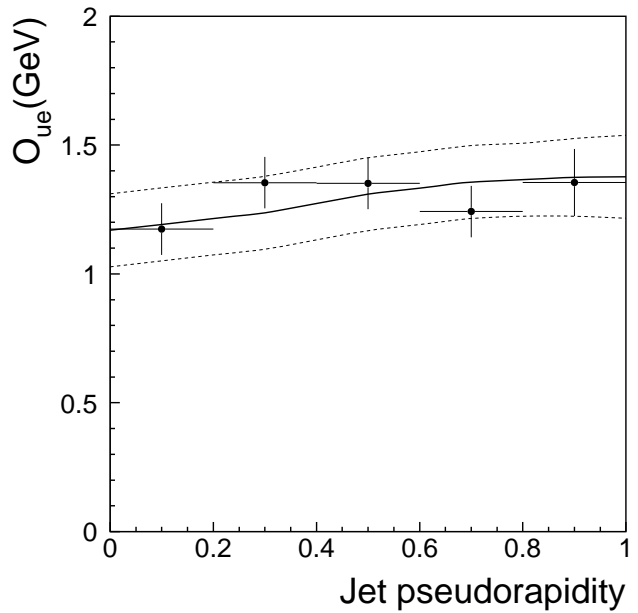}
\vspace{0.2cm}
\caption{
The correction for underlying event $O_{{\rm ue}}$ 
as a function of $|\eta|$ for $k_{\perp}$ jets ($D = 1.0$).
The solid curve is the fit of the 
results for the cone jet algorithm in Ref.~\protect\cite{jes} scaled to the
results for the $k_{\perp}$ jet algorithm.
The dashed curves denote the one standard deviation 
(s.d.) systematic error.
}
\label{fig:ue_new}
\end{figure}

The offset due to the underlying event $O_{{\rm ue}}$ is modeled with
minimum-bias events.  A minimum-bias event is a zero-bias event
with the additional requirement of a coincidence signal
in the scintillating-tile hodoscopes~\cite{d0} near the beampipe.
The additional requirement means there was an inelastic
$p\bar{p}$ collision during the
bunch crossing.
In addition to $O_{{\rm ue}}$, a minimum-bias event in the D\O\ calorimeter
includes energy from uranium noise, multiple interactions, and pile-up.
However, the luminosity dependence of multiple interactions and pile-up
in minimum-bias events is different 
than in zero-bias events.
In the limit of very small luminosity, these contributions are negligible,
and a minimum-bias event at low luminosity therefore contains
the offset due to the underlying event and uranium noise,
while
a zero-bias event at low luminosity has only
the offset from uranium noise.
To measure $O_{{\rm ue}}$,
we again compare two samples of jets.
Minimum-bias events as measured by the D\O\ calorimeter at low luminosity
are added to Monte Carlo jet events, where the resulting jets define the first sample
of jets in the determination of $O_{{\rm ue}}$.
The second sample of jets is reconstructed from
zero-bias events at low luminosity
and also added to Monte Carlo jet events. 
On an event-by-event basis,
$O_{{\rm ue}}$ is calculated by subtracting the
momentum of jets in the second sample 
from the momentum of matching jets in the first sample.
The underlying event offset $O_{{\rm ue}}$ for $k_{\perp}$ jets is
shown in Fig.~\ref{fig:ue_new}.
Using this method for both algorithms, the offset 
$O_{{\rm ue}}$ for $k_{\perp}$ jets (with $D = 1.0$) is 
found to be approximately $30\%$ 
larger than for the fixed-cone jet algorithm 
(with ${\cal R}  = 0.7$).

D\O\ measures the jet momentum response 
based on
conservation of $p_T$ in photon-jet ($\gamma$-jet)
events~\cite{jes}. 
The electromagnetic energy/momentum scale is
determined from the
$Z, J/\psi \rightarrow e^{+}e^{-}$, and
$\pi^{0} \rightarrow \gamma \gamma \rightarrow e^{+}e^{-}e^{+}e^{-}$ 
data samples, using the known masses of these
particles.
For the case of a $\gamma$-jet two-body process, the jet momentum response
can be characterized as:
\begin{equation} 
 R_{{\rm jet}} =
  1 + \frac{\vmet \cdot \hat{n}_{T\gamma}}{p_{T\gamma}} \, ,
\label{eq:mpf}
\end{equation}
where $p_{T\gamma}$ and $\hat{n}$ are the transverse
momentum and direction of the photon,
and \vmet\ is the missing transverse energy,
defined as the negative of the vector sum
of the transverse energies of the cells in the calorimeter. 
To avoid resolution
and trigger biases, $R_{{\rm jet}}$ is binned in terms of $E^{\prime} =
p_{T\gamma}^{{\rm meas}}
\cdot {\rm cosh} (\eta_{{\rm jet}})$ and then mapped onto
the offset-corrected jet momentum.
Thus, 
$E^{\prime}$ depends only on photon variables and
jet pseudorapidity, which are quantities that are measured with very good
resolution.  $R_{{\rm jet}}$ and $E^{\prime}$ depend only on the jet
position, which has little dependence on the type of jet algorithm
employed.

\begin{figure}
\epsfxsize=3.375in
\epsfbox{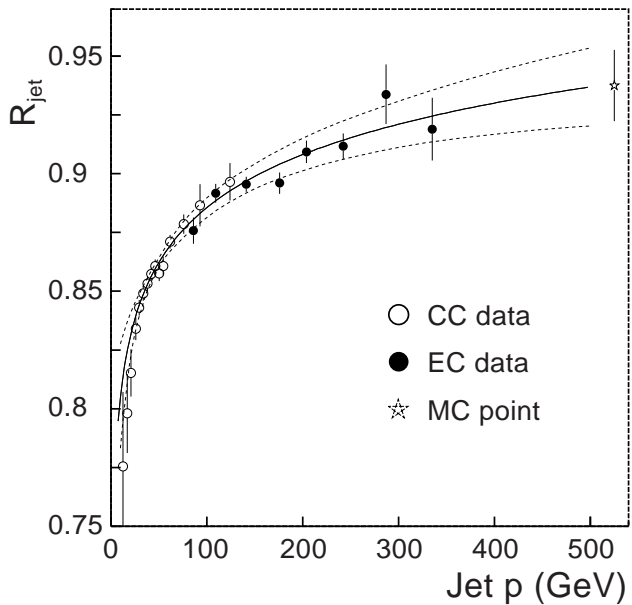}
\vspace{0.2cm}
\caption{The response correction for $k_{\perp}$
jets with $D = 1.0$, as a function of offset-corrected jet momentum.
The Monte Carlo point ($\star$) is used to constrain the fit (solid)
at high $p_{{\rm jet}}^{{\rm meas}}$.
The dashed curves denote the $\pm 1$ s.d. systematic error.
}
\label{fig:kt_resp}
\end{figure}

$R_{{\rm jet}}$ as a function of $p^{{\rm meas}}_{{\rm jet}} - p_O$ 
for $k_{\perp}$ jets
is shown in Fig.~\ref{fig:kt_resp}.  The data points are
fitted with the functional form $R_{{\rm jet}}(p) = a + b \: {\rm ln}(p)
+ c \: ({\rm ln}(p))^2$. 
The response $R_{{\rm jet}}$ for cone jets (with ${\cal R}=0.7$)~\cite{jes} 
and for $k_{\perp}$ jets ($D = 1.0$)
is different by about 0.05. This difference does not have any 
physical meaning; it corresponds to different voltage-to-energy conversion
factors at the cell level used in the reconstruction of jets.

\subsection{Comparison of the $\lowercase{k}_{\perp}$ jet algorithm to the cone jet algorithm}
\label{subs:comp}

It is of interest to compare the momenta of $k_{\perp}$ jets 
to those of jets reconstructed with the
D\O\ fixed-cone algorithm~\cite{run2qcd}.
These results refer to the $k_{\perp}$ jet algorithm described above
with $D = 1.0$ and corrected according to 
the prescriptions given in Sec.~\ref{subs:jes}.
The cone jets were reconstructed~\cite{iain} with ${\cal R} = 0.7$ 
and corrected according to Ref.~\cite{jes}.
This comparison involves about $75\%$ of 
the events in the 1994--1996 data
that were used for the analysis of the inclusive cone-jet cross section 
at $\sqrt{s} = 1800$ GeV~\cite{inc}.
The two algorithms are similar by design~\cite{ES},
defining similar jet directions and momenta,
at least for the two leading (highest $p_T$) jets in the event.
The remaining jets in the event usually have much smaller $p_T$,
making them more difficult to measure, and so we do not consider them here.
The jets reconstructed by each algorithm are compared on an
event-by-event basis, associating a cone jet with a $k_{\perp}$ jet
if they are separated by $\Delta{\cal R} < 0.5$.

\begin{figure}
\epsfxsize=3.375in
\epsfbox{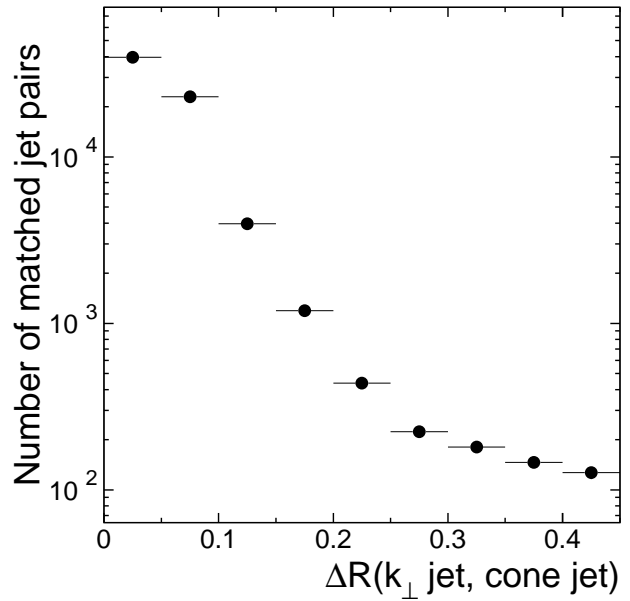}
\vspace{0.2cm}
\caption
{
The distance $\Delta{\cal R} = \sqrt{ \Delta \eta^2 + \Delta \phi^2}$
between a $k_{\perp}$-jet axis and its matching cone-jet axis.
The $k_{\perp}$ jets were reconstructed with $D = 1.0$,
and the cone jets were reconstructed with ${\cal R} = 0.7$.
Only the two leading jets from each algorithm were considered.
The $k_{\perp}$ jets were selected with $|\eta| < 0.5$.
}
\label{fig:kt_cone_delta_r}
\end{figure}

To obtain a sample of events with only good hadronic jets,
the following requirements were placed on the 
events and on the leading two reconstructed $k_{\perp}$ jets.
These criteria are based 
on standard jet quality requirements (to remove spurious clusters) 
in use at D\O\ for the fixed-cone jet algorithm~\cite{iain}:
\begin{itemize}
\item Measured event vertex was required to be 
within 50 cm of the center of the detector.
\item $|\vmet|$ was required to be less than $70\%$ of the $p_T$ of the leading jet.
\item Fraction of jet $p_T$ measured in the coarse hadronic
calorimetry was required to be less than $40\%$ of the total jet $p_T$.
\item Fraction of jet $p_T$ measured in the electromagnetic
calorimetry was required to be between $5\%$ and $95\%$ of the total jet $p_T$.
\item Jets were required to have $|\eta| < 0.5$.
\end{itemize}

\begin{figure}
\epsfxsize=3.375in
\epsfbox{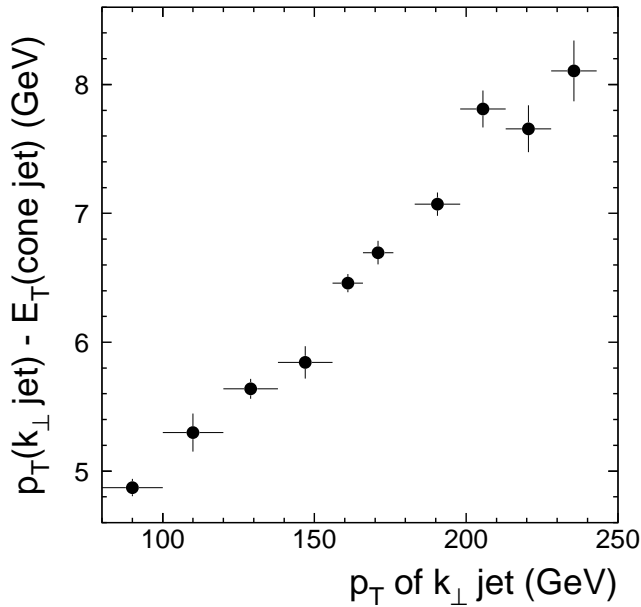}
\vspace{0.2cm}
\caption
{
The difference 
$p_T(k_{\perp} \text{ jet}) - E_T(\text{cone jet})$ 
as a function of the $k_{\perp}$ jet $p_T$.
A cone jet is associated with a $k_{\perp}$ jet
if their axes are separated by $\Delta{\cal R}  < 0.5$.
}
\label{fig:kt_cone_delta_et}
\end{figure}

These requirements yield a sample of 
68946
$k_{\perp}$ jets.
The axes of $99.94\%$ of these jets
are reconstructed within $\Delta{\cal R} < 0.5$ of a cone-jet axis,
when the matching jet is one of the two leading cone jets in the event.
For such pairs of jets, 
the distance between a $k_{\perp}$-jet axis and matching cone-jet axis is shown
in Fig.~\ref{fig:kt_cone_delta_r}.
The fixed-cone algorithm finds a jet within  
$\Delta{\cal R} < 0.1$ of a $k_{\perp}$ jet $91\%$ of the time.
Figure~\ref{fig:kt_cone_delta_et} shows the difference 
$p_T(k_{\perp} \, {\rm jet}) - E_T({\rm cone \, jet})$ 
as a function of $p_T(k_{\perp} \, \text{jet})$.
Generally, the $p_T$ of 
$k_{\perp}$ jets ($D = 1.0$) is higher than the 
$E_T$ of associated cone jets (${\cal R} = 0.7$).
The difference increases approximately linearly with jet $p_T$,
from about 5 GeV (or $6 \%$) at $p_T \approx 90$ GeV
to about 8 GeV (or $3 \%$) at $p_T \approx 240$ GeV.
This may be explained by how the two algorithms
deal with hadronization effects~\cite{seb}.

\subsection{Subjets}
\label{subs:sub}
The subjet multiplicity is a natural observable
for characterizing a $k_{\perp}$ jet~\cite{cat93,cat92}.
Subjets are defined by reapplying the $k_{\perp}$ algorithm,
as in Sec.~\ref{subs:clu}, 
starting with a list of 
preclusters assigned to 
a particular jet.
Pairs of objects with the smallest $d_{ij}$ are merged successively until
all remaining pairs of objects have
\begin{equation}
d_{ij} =  \min(p_{T,i}^2,p_{T,j}^2) \frac{ \Delta{\cal R}_{ij}^2} {D^2} > y_{{\rm cut}} p_T^{2}({\rm jet})\label{eq:y}, 
\end{equation}
where $p_T({\rm jet})$ is the $p_T$ of the entire jet
in the $k_{\perp}$ algorithm described above, and
$0 \le y_{{\rm cut}} \le 1$ is a dimensionless parameter.
Objects satisfying Eq.~(\ref{eq:y})
are called subjets, and the number of subjets
is the subjet multiplicity $M$ of a $k_{\perp}$ jet.
For $y_{{\rm cut}} = 1$, the entire jet consists of a single subjet ($M=1$).
As $y_{{\rm cut}}$ decreases, the subjet multiplicity increases, until
every precluster becomes resolved as a separate subjet
in the limit $y_{{\rm cut}} \rightarrow 0$.
Two subjets in a jet can be
resolved when they are not collinear 
(i.e., well-separated in $\eta \times \phi$ space),
or if they are both hard (i.e., carry a significant fraction
of the jet $p_T$).

We now turn to the theoretical treatment of subjet multiplicity.
Perturbative and resummed calculations~\cite{sey_sub1,sey_sub2}
and Monte Carlo estimates (see Sec.~\ref{subs:ext})
predict that gluon jets have a higher mean subjet multiplicity
than quark jets.
To understand the origin of this prediction,
we
consider first how a jet can contain multiple subjets.
Clearly, at
leading-order, $2 \rightarrow 2$ subprocesses
yield
$M = 1$.
However, higher-order QCD radiation can
increase the average value of $M$.
At next-to-leading order, there can be
three partons in the final state of a
$p\bar{p}$ collision.
If two partons are clustered together into a jet,
they can be resolved 
as 
distinct
subjets ($M = 2$)
for a sufficiently small choice of $y_{{\rm cut}}$.
For larger $y_{{\rm cut}}$, the value of $M$ depends on
the magnitude and direction of the radiated third parton.
In QCD, the radiation of a parton
is governed by the DGLAP splitting functions~\cite{dglap}.
The radiated third parton is usually soft and/or collinear
with one of the other two partons, leading to jets with
$M = 1$.
However, hard or large-angle radiation, although rare,
causes
some jets to have $M = 2$.
Consequently, when many jets are analyzed using some high $y_{{\rm cut}}$,
the two-subjet rate
will yield $\langle M \rangle > 1$.

In the framework of parton showers, repeated application of DGLAP
splitting provides jets with $M > 2$.
Monte Carlo event generators incorporate
parton showers into the initial and final states 
of a $2 \rightarrow 2$ hard scatter.
Because of its larger color factor,
a parton shower initiated by a gluon in the final state
will tend to produce a jet with more subjets
than one initiated by a quark.
Similarly, a soft parton radiated in the initial state
will tend to cluster with a hard final-state parton 
when $\Delta{\cal R} < D$.
For the case of initial-state radiation, 
the subjet multiplicity depends weakly on whether 
the final-state partons in the $2 \rightarrow 2$ 
hard scatter are quarks or gluons.
The contribution of initial-state radiation to the subjet multiplicity
does, however, depend on $\sqrt{s}$.
Initial-state radiation is treated on an equal footing 
as final-state radiation in the 
$k_{\perp}$ algorithm with $D = 1.0$~\cite{sey_sub1,sey_pbarp},
and diminishes in importance as $D$ decreases.
In general, subsequent emissions in parton showers
have less energy and momentum, 
and this structure is revealed at smaller $y_{{\rm cut}}$ values
through an increase in the subjet multiplicity:
$\langle M(y_{{\rm cut}}') \rangle >\langle M(y_{{\rm cut}}) \rangle$,
where $y_{{\rm cut}}' < y_{{\rm cut}}$.

Experimentally, the growth of $M$ at very small $y_{{\rm cut}}$
is reduced by the granularity
of the detector and by the preclustering algorithm.  
Theoretical predictions for $M$ are therefore 
treated in the same way as the
experimental measurements, i.e., by preclustering (as in Sec.~\ref{subs:pre}).
Requiring preclusters to be separated by 
$\Delta{\cal R}^{{\rm pre}}$, means that
the subjets nearest in 
$(\eta,\phi)$ space begin to be resolved for 
\begin{equation}
y_{{\rm cut}} < 
\left( \frac{\Delta{\cal R}^{{\rm pre}}} {2D} \right)^2
\end{equation}
based solely on the fraction of $p_T$ carried by the subjet in the jet.
The factor 1/2 corresponds to the maximum fraction of jet $p_T$
carried by the softest subjet [see Eq.~(\ref{eq:y})].
The preclustering stage provides a comparison of the measurement of
$M$ with prediction in the interesting region of small $y_{{\rm cut}}$,
without an explicit correction for detector granularity.

The subjet analysis in this paper uses a single resolution 
parameter $y_{{\rm cut}} = 10^{-3}$.  
For this $y_{{\rm cut}}$, the minimum subjet $p_T$
is approximately $3\%$ of the total jet $p_T$, 
independent of the choice of the $D$ parameter.
Because $y_{{\rm cut}}$, as defined by Eqs.~(\ref{eq:dij}) and (\ref{eq:y}),
involves a ratio of subjet $p_T$ to jet $p_T$,
the subjet multiplicity is therefore not significantly sensitive
to multiplicative changes in the overall 
$p_T$ scale.
Consequently, given the fact that subjets are specified 
during jet reconstruction,
and the jet momentum calibration is derived after reconstruction,
we do not attempt to correct the momenta of individual subjets.
However, the subjet multiplicity
is corrected for the experimental effects that cause an offset
in jet $p_T$.
In general, the presence of uranium noise, multiple interactions, and pile-up,
tends to increases the subjet multiplicity.

\section{Data samples}
\label{subs:sel}
In leading-order QCD, 
the fraction of final-state jets originating from gluons decreases
with increasing $x \propto p_T / \sqrt{s}$, 
the momentum fraction carried by the initial-state partons.
This is due primarily to the $x$-dependence of the 
parton distributions.
Because, for fixed $p_T$, the gluon fraction decreases when 
$\sqrt{s}$ is decreased from 1800 GeV to 630 GeV,
this suggests an experimental way to define jet samples with
different mixtures of quarks and gluons.
A single set of criteria can be used to select jets at the two
beam energies, without changing any of the detector elements.
We use this principle to analyze an event sample recorded at the end of 1995 
by the D\O\ detector at $\sqrt{s} = 630$ GeV,
and compare it with the larger 1994--1995 event sample 
collected at $\sqrt{s} = 1800$ GeV.
The lower range of jet $p_T$ populated by the smaller event
sample at $\sqrt{s} = 630$ GeV dictated
the ultimate criteria used in the comparison.
In Sec.~\ref{subs:lo}, we first
describe a simple test of
a set of criteria used to select quark-enriched 
and gluon-enriched jet samples.
In Sec.~\ref{subs:seld0}, we specify each criterion 
used in the analysis.
In Sec.~\ref{subs:selmc}, we provide a Monte Carlo estimate of
the quark/gluon yield based on the full set of criteria.
Finally, in Sec.~\ref{subs:ext}, we describe how to 
estimate the subjet content
of gluon and quark jets.

\subsection{Gluon and quark samples at leading-order in QCD}
\label{subs:lo}
\begin{figure} [t]
\epsfxsize=3.375in
\epsfbox{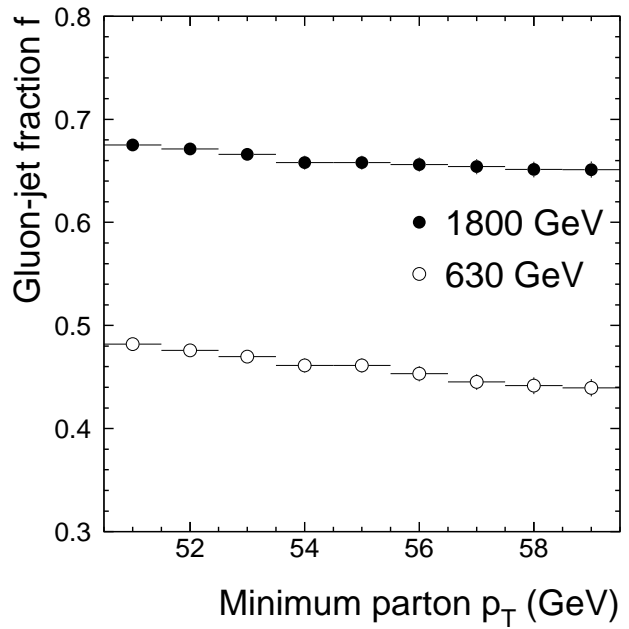}
\vspace{0.2cm}
\caption{The Monte Carlo gluon-jet fraction $f$ at leading-order,
for final-state partons with 
maximum parton $p_T = 100$ GeV,
and
minimum parton
$p_T \approx 55$ GeV,
as a function of 
the minimum parton
$p_{T}$, 
using the CTEQ4M PDF.
Both partons are required to be central 
($|\eta| < 0.5$).
The solid symbols show the
prediction for $\sqrt {s}=$ 1800 GeV,
and the open symbols show the
prediction for $\sqrt {s}=$ 630 GeV.
}
\label{fig:isid34_cut_integ_100_big}
\end{figure}

For a given set of 
parton distribution functions (PDFs),
the relative admixture of gluon and quark jets passing
a set of kinematic criteria can be estimated using a
leading-order QCD event generator.
At this order, there is no dependence on jet algorithm,
because each of the two final-state partons defines a jet.
We use the {\sc HERWIG} v5.9 Monte Carlo with the CTEQ4M~\cite{cteq}
PDFs
to generate
leading-order QCD $2 \rightarrow 2$ events,
and keep track of the identity (gluon or quark) of the 
partons.
At leading order, the gluon-jet fraction $f$
corresponds to the
number of final-state gluons that pass the selections
divided by the total 
number of final-state partons that pass the selections.
For example, the jet sample selected from only $gg \rightarrow gg$
or $q\bar{q} \rightarrow gg$
events will have a gluon-jet fraction of unity.
Figure~\ref{fig:isid34_cut_integ_100_big}
shows that for
the full ensemble of Monte Carlo events,
the gluon-jet fraction
at $\sqrt{s} = 630$ GeV 
is about $30\%$ smaller
than at $\sqrt{s} = 1800$ GeV,
where we have selected central 
($|\eta| < 0.5$)
jets with 
minimum parton
$p_T \approx 55$ GeV
and maximum parton $p_T = 100$ GeV.
This difference is due primarily to the 
relative abundance of initial-state gluons at these $x$ values for
$\sqrt{s} = 1800$ GeV compared to
$\sqrt{s} = 630$ GeV.

\subsection{Jet data samples}
\label{subs:seld0}
We define gluon-enriched and 
quark-enriched central ($|\eta| < 0.5$) jet samples 
using identical criteria at $\sqrt{s} = 1800$ GeV and 630 GeV,
thereby reducing any experimental biases and systematic effects.
We select events that pass a trigger 
requiring the scalar sum of $E_T$ above 30 GeV 
within a cone of size
${\cal R} = 0.7$ 
~\cite{iain},
and apply the selections listed in Sec.~\ref{subs:comp},
but only for jets with measured $p_T$ between 55 and 100 GeV.
These cuts yield samples of 11,007 jets at $\sqrt{s} = 1800$ GeV,
and 1194 jets at $\sqrt{s} = 630$ GeV.

\begin{figure}
\epsfxsize=3.375in
\epsfbox{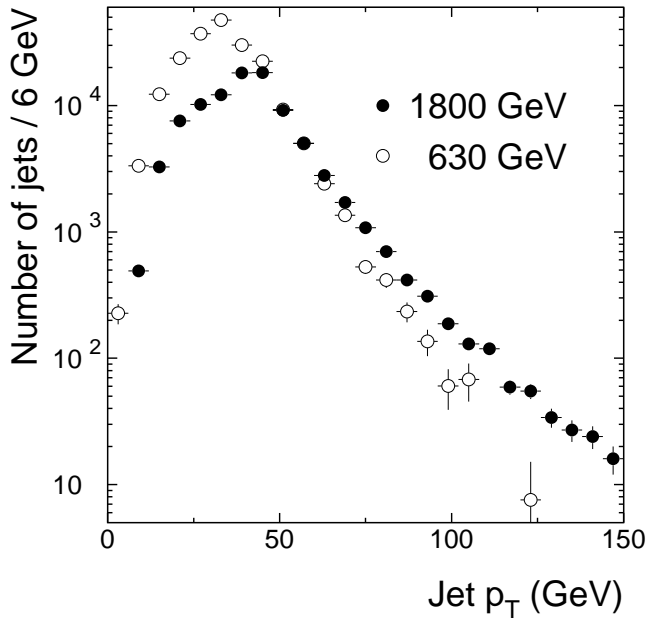}
\vspace{0.2cm}
\caption{
The $p_{T}$ distribution of selected central ($|\eta| < 0.5$) 
jets in D\O\ data, before applying a cutoff on jet $p_T$.
The data at $\sqrt {s}=$ 630 GeV are normalized to the data at
$\sqrt {s}=$ 1800 GeV in the bin $54 \le p_T < 60$ GeV.
The turnover at lower jet $p_T$ is due to inefficiencies in the
trigger.
For the following analysis, we use jets with $55 < p_T < 100$ GeV.
}
\label{fig:kt_et}
\end{figure}

An important point is that these jets were 
reconstructed with the $k_{\perp}$ algorithm for $D = 0.5$.
This choice tends to select events with fewer subjets from 
initial-state radiation,
which can vary with $\sqrt{s}$ (see Sec.~\ref{subs:sub}).
Figure~\ref{fig:kt_et} shows that the $p_T$ distribution
of the selected jets at $\sqrt{s} = 1800$ GeV is
harder than at $\sqrt{s} = 630$ GeV.
The mean jet $p_T$ at $\sqrt{s} = 1800$ GeV is
$66.3 \pm 0.1$ GeV, which is 
2.3 GeV higher than at $\sqrt{s} = 630$ GeV.
This cannot be caused by any differences in the contribution
to the offset in the jet $p_T$.
In fact, the entire offset is $p_O \approx 3-4$ GeV per jet 
at $\sqrt{s} = 1800$ GeV
for $D = 1.0$ (see Sec.~\ref{subs:jes}),
and is therefore an expected factor $\approx \! 4$ smaller for $D = 0.5$.
Moreover, only a small fraction of the jet offset 
can be attributed to the difference in $\sqrt{s}$.
Even so, offset differences can only change 
the subjet multiplicity by shifting
the relative jet $p_T$.
Rather than attempting to measure and account for such small effects
in the jet $p_T$ distributions, we simply use identical
jet criteria at the two beam energies,
and estimate the uncertainty on $M$ by varying the jet selection cutoffs
(see Sec.~\ref{subs:sys}).

\subsection{Jet samples in Monte Carlo events}
\label{subs:selmc}
\begin{figure}
\centerline{\psfig{figure=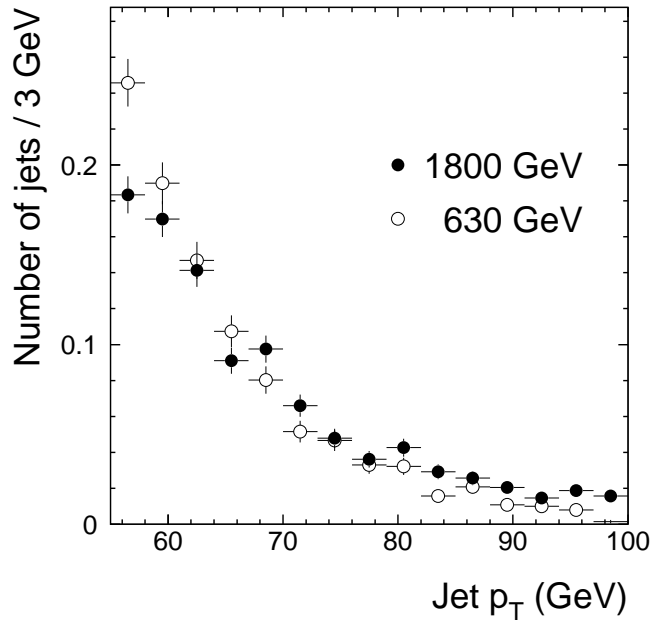,width=8.6cm}}
\caption{The normalized $p_{T}$ distribution of central ($|\eta| < 0.5$) 
jets selected in Monte Carlo events at $\sqrt {s}=$ 1800 GeV and 630 GeV.
Each distribution
has been normalized to unit area.
}
\label{fig:uz_c47_11}
\end{figure}

To estimate the number of gluon jets in the $\sqrt{s} = 1800$ 
GeV and 630 GeV jet samples,
we generated approximately 10,000 {\sc HERWIG} events at each $\sqrt{s}$, 
with parton $p_T > 50$ GeV, and requiring at least one of the
two leading-order partons to be central ($|\eta| < 0.9$).
The events were passed through a full simulation of the 
D\O\ detector.  To simulate the effects of uranium noise,
pile-up from previous bunch crossings, and multiple $p\bar{p}$
interactions in the same bunch crossing,
we overlaid D\O\ random-crossing events onto our Monte Carlo
sample, on a cell-by-cell basis in the calorimeter.
(A sample with instantaneous luminosity of 
${\cal L} \approx 5 \times 10^{30} \text{cm}^{-2}\text{s}^{-1}$
was used at $\sqrt{s} = 1800$ GeV, and
${\cal L} \approx 0.1 \times 10^{30} \text{cm}^{-2}\text{s}^{-1}$
was used at $\sqrt{s} = 630$ GeV.)
These pseudo events were then passed through the normal 
offline-reconstruction
and jet-finding packages.
Jets were then selected using the same criteria as used for D\O\ data,
and their $p_T$ distribution 
is shown in Fig.~\ref{fig:uz_c47_11}.

\begin{figure} 
\epsfxsize=3.375in
\epsfbox{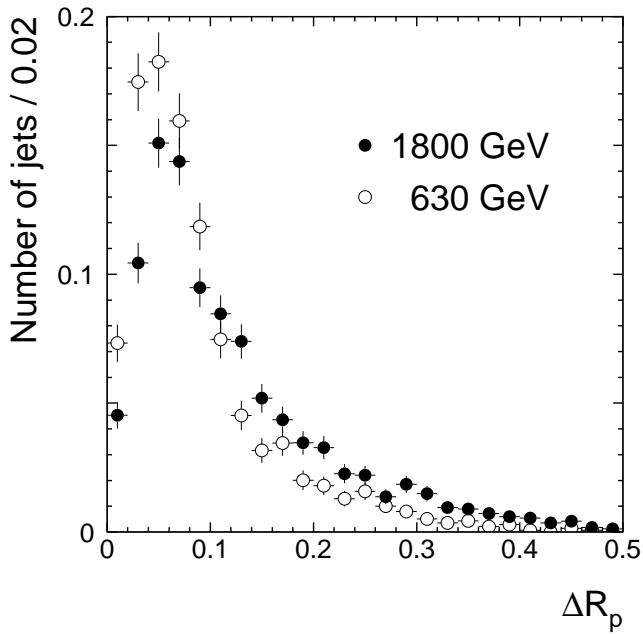}
\vspace{0.2cm}
\caption{
The distance of the closest
calorimeter-level Monte Carlo jet to one of the leading final-state partons. 
The solid (open) points show the Monte Carlo 
sample at $\sqrt{s}=1800$ (630) GeV.
Each distribution
has been normalized to unit area.
}
\label{fig:mc_dr}
\end{figure}

\begin{figure} 
\epsfxsize=3.375in
\epsfbox{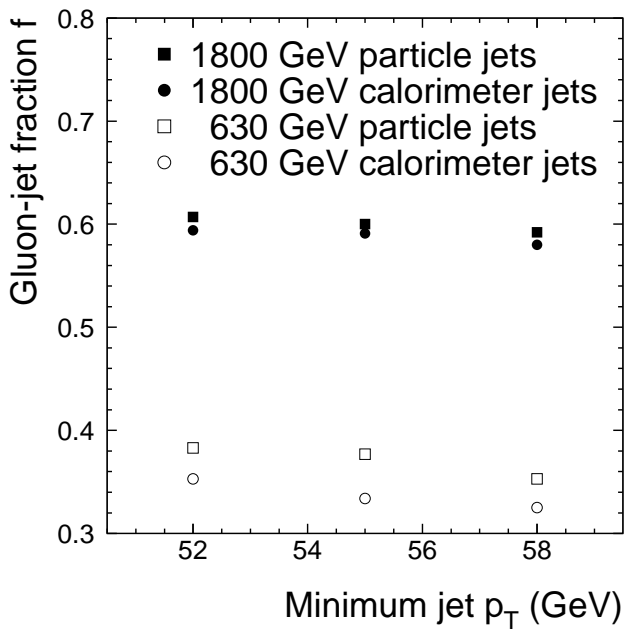}
\vspace{0.2cm}
\caption{The gluon-jet fraction of selected jets with 
maximum $p_{T} = 100$ GeV
and
minimum $p_{T}$ between 52 and 58 GeV,
as a function of 
minimum jet $p_{T}$,
for $\sqrt {s}=$ 1800 GeV and 630 GeV,
using the CTEQ4M PDF.
The jets have been tagged through the identity 
of the nearer leading-order final-state parton.
}
\label{fig:gluon_frac_vs_etmin}
\end{figure}

We tag each such selected Monte Carlo jet
as either quark or gluon based on the identity of the nearer 
(in $\eta \times \phi$ space)
final-state parton in the QCD $2 \rightarrow 2$ hard scatter.
The distance between one of the
partons and the closest calorimeter jet is shown in 
Fig.~\ref{fig:mc_dr}.
There is clear correlation between jets in the calorimeter
and partons from the hard scatter.
The fraction of gluon jets is shown in
Fig.~\ref{fig:gluon_frac_vs_etmin}
as a function of the minimum $p_T$ used to select the jets.
There is good agreement for the gluon-jet fraction
obtained using
jets reconstructed at the calorimeter
and at the particle levels ($\Delta f  < 0.03$).
The smaller gluon-jet fractions relative to
leading-order
(Fig.~\ref{fig:isid34_cut_integ_100_big})
are due mainly to the presence of 
higher-order radiation in the QCD Monte Carlo.
When $p_T$ cutoffs are applied to particle-level jets, the associated
leading-order partons shift to significantly higher $p_T$.
Since the gluon-jet
fraction decreases with increasing parton $p_{T}$, $f$ is
smaller when events are selected according to particle-level jet $p_{T}$
rather than when they are selected according to partonic $p_{T}$.
The same is true for cutoffs applied to the calorimeter-level jets 
compared to the particle-level jets,
although here the $\Delta f$ discrepancy is much smaller.
In what follows, we shall use nominal
gluon-jet fractions $f_{1800}=0.59$ 
and $f_{630}=0.33$, obtained from Monte Carlo at the calorimeter level
for $55< p_{T} < 100$ GeV.

\subsection{Subjets in gluon and quark jets}
\label{subs:ext}
Using the previously described jet samples,
there is a simple way to distinguish between
gluon and quark jets on a statistical basis~\cite{Acton:1993jm}.
The subjet multiplicity 
in a mixed sample of gluon and quark jets
can be written as a linear combination of subjet multiplicity
in gluon $M_g$ and quark jets $M_q$:
\begin{equation}
M=fM_{g}+(1-f)M_{q}
\label{eq:m}
\end{equation}
The coefficients are the fractions of gluon and
quark jets in the mixed sample, $f$ and $(1-f)$, respectively. 
Considering Eq.~(\ref{eq:m}) for two
samples of jets at $\sqrt{s} = 1800$ GeV and 630 GeV,
and assuming that $M_{g}$ and $M_{q}$
are independent of $\sqrt{s}$ (we address this assumption later),
we can write:
\begin{equation}
M_{g}=\frac{\left( 1-f_{630}\right) M_{1800}-\left( 1-f_{1800}\right) M_{630}%
}{f_{1800}-f_{630}}  \label{eq:mg}
\end{equation}
\begin{equation}
M_{q}=\frac{f_{1800}M_{630}-f_{630}M_{1800}}{f_{1800}-f_{630}}  \label{eq:mq}
\end{equation}
where $M_{1800}$ and $M_{630}$ are the
measured multiplicities in the mixed-jet samples at  
$\sqrt{s} = 1800$ GeV and 630 GeV,
and $f_{1800}$ and $f_{630}$ are the gluon-jet fractions in the two samples.
The extraction of $M_{g}$ and $M_{q}$ requires prior
knowledge of the two gluon-jet fractions,
as described in Sec.~\ref{subs:selmc}.
Since the gluon-jet fractions depend on jet $p_T$ and $\eta$,
Eqs.~(\ref{eq:mg}) and (\ref{eq:mq}) hold only within
restricted regions of phase space, 
i.e., over small ranges of jet $p_T$ and $\eta$.
Equations~(\ref{eq:mg}) and (\ref{eq:mq}) can, of course, 
be generalized to any observable
associated with a jet.  

\begin{figure}
\epsfxsize=3.375in
\epsfbox{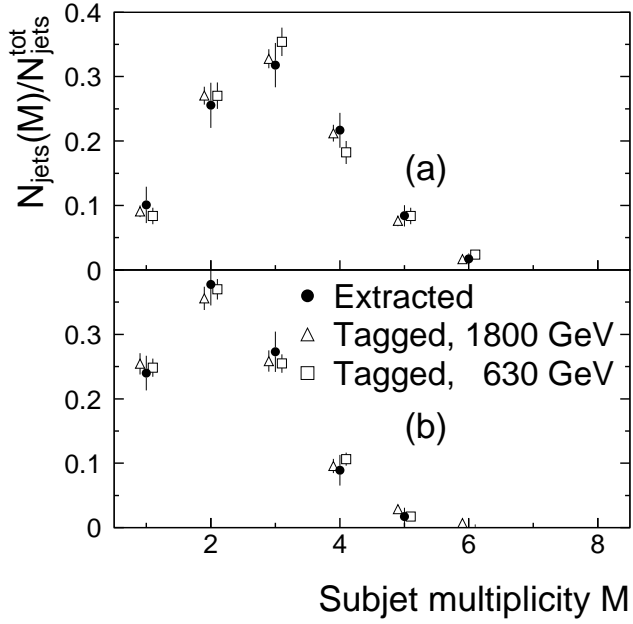}
\vspace{0.2cm}
\caption{Uncorrected subjet multiplicity in 
fully-simulated Monte Carlo (a) gluon and (b) quark
jets. 
The number of jets $N_{{\rm jets}}(M)$ in each bin of
subjet multiplicity on the vertical axis is normalized
to the total number of jets in each sample 
$N_{{\rm jets}}^{{\rm tot}} = \sum_{M}  N_{{\rm jets}}(M)$.
The measured distributions (solid) are extracted from the
mixed Monte Carlo jet samples at $\sqrt{s} = 1800$ GeV and 630 GeV.
The tagged distributions (open) 
are for $\sqrt{s} = 1800$ GeV (triangles) 
and 630 GeV (squares).
}
\label{fig:mc}
\end{figure}

We use our Monte Carlo samples to check
Eqs.~(\ref{eq:mg}) and (\ref{eq:mq})
for $k_{\perp}$ jets reconstructed using the full-detector
simulation with $D = 0.5$.
Such a consistency test does not depend on the 
details of the subjet multiplicity
distributions ($M_q, M_g, M_{1800}, M_{630}$).
The extracted
distributions in $M_g$ and $M_q$
are shown in Fig.~\ref{fig:mc}.
As expected, Monte Carlo gluon jets 
have more subjets, 
on average, 
than Monte Carlo quark jets: 
$\langle M_{g} \rangle > 
\langle M_{q} \rangle$.
This is also found for jets reconstructed at the particle level,
and the differences between gluon and quark jets do not appear
to be affected by the detector.
Also, the subjet multiplicity distributions for tagged jets 
are similar at the two center-of-mass energies,
verifying the assumptions used in deriving 
Eqs.~(\ref{eq:mg}) and (\ref{eq:mq}).
Finally, the extracted $M_q$ and $M_g$ distributions
agree very well with the tagged distributions.
This demonstrates self-consistency of the extraction using
Eqs.~(\ref{eq:mg}) and (\ref{eq:mq}).

\section{Subjet multiplicities}
\subsection{Uncorrected subjet multiplicity}
\label{subs:raw}
\begin{figure}
\epsfxsize=3.375in
\epsfbox{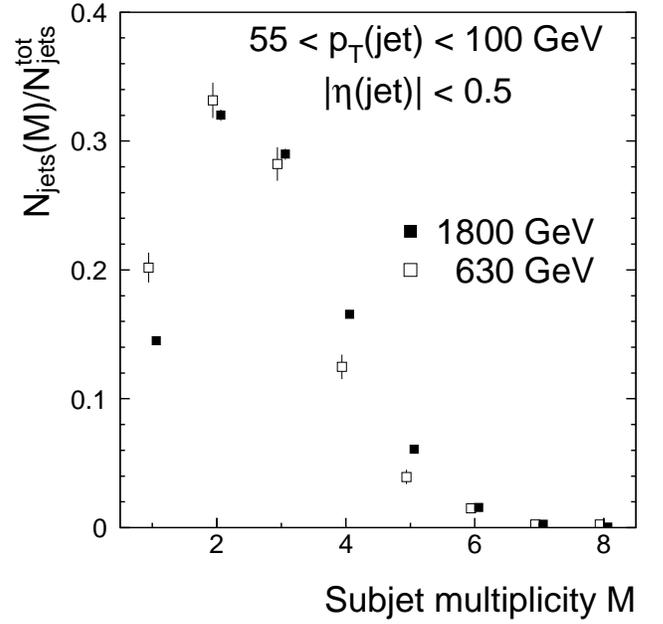}
\vspace{0.2cm}
\caption{
Uncorrected subjet multiplicity in jets from D\O\ data at 
$\sqrt{s} = 1800$ GeV and 630 GeV.
}
\label{fig:1800_630}
\end{figure}

\begin{figure}
\epsfxsize=3.375in
\epsfbox{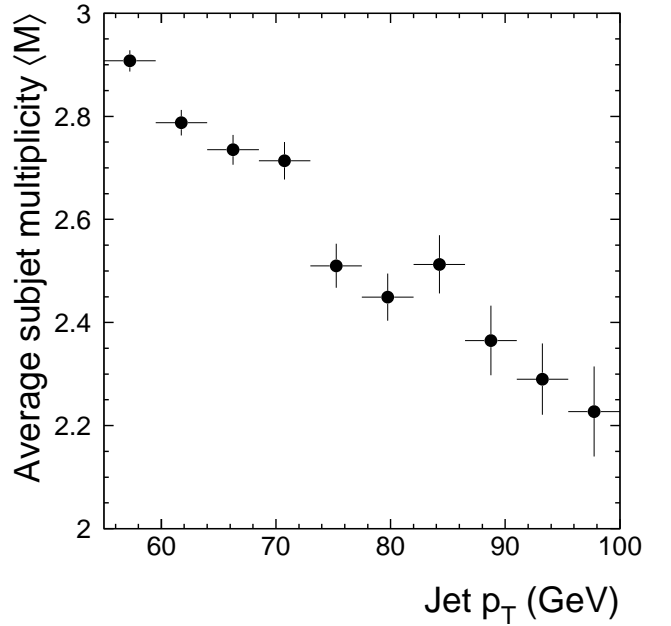}
\vspace{0.2cm}
\caption{
Uncorrected mean subjet multiplicity versus jet $p_T$
in D\O\ data at 
$\sqrt{s} = 1800$ GeV.
Note the suppressed zero on the vertical axis.
}
\label{fig:nsub_vs_et}
\end{figure}

Figure~\ref{fig:1800_630} shows the distributions
of subjet multiplicity
for the D\O\ data samples described in Sec.~\ref{subs:sel}.
This is the first measurement of its kind at a hadron collider.
The average number of subjets in jets at $\sqrt{s} = 1800$ GeV
is $\langle M_{1800} \rangle = 2.74 \pm 0.01$, where the error
is statistical.  This is higher than the value of
$\langle M_{630} \rangle = 2.54 \pm 0.03$
at $\sqrt{s} = 630$ GeV.
The observed shift is consistent with the prediction
that there are more gluon jets in the sample at $\sqrt{s} = 1800$ GeV 
than in the sample at 
$\sqrt{s} = 630$ GeV, and that gluons radiate more subjets than quarks do.
The fact that the $p_T$ spectrum is harder at $\sqrt{s} = 1800$ GeV
than at $\sqrt{s} = 630$ GeV cannot be the cause of this effect
because the subjet multiplicity decreases with increasing jet $p_T$.
Figure~\ref{fig:nsub_vs_et} shows the rather mild dependence of the
average subjet multiplicity on
jet $p_T$.

\begin{figure}
\epsfxsize=3.375in
\epsfbox{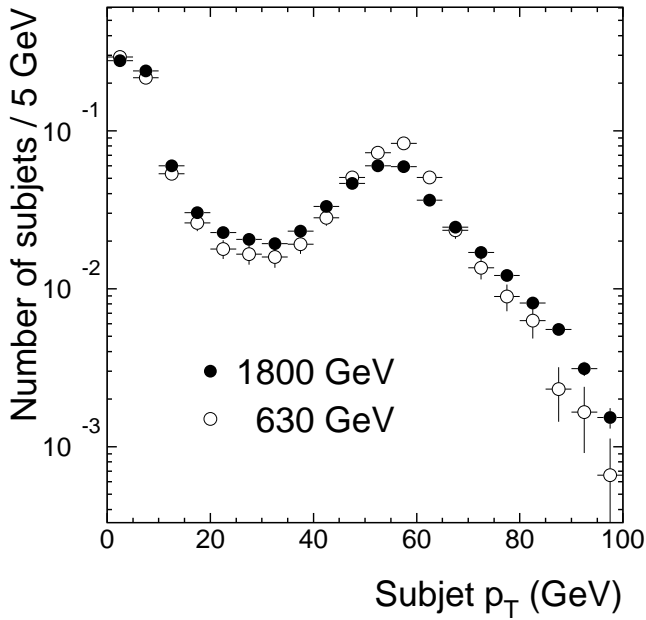}
\vspace{0.2cm}
\caption{
The uncorrected $p_{T}$ distribution of subjets in 
data for jets with $55 < p_T < 100$ GeV
and $|\eta| < 0.5$.
All selections have been applied,
and each distribution
has been normalized to unit area.
}
\label{fig:sub_et}
\end{figure}

\begin{figure}
\epsfxsize=3.375in
\epsfbox{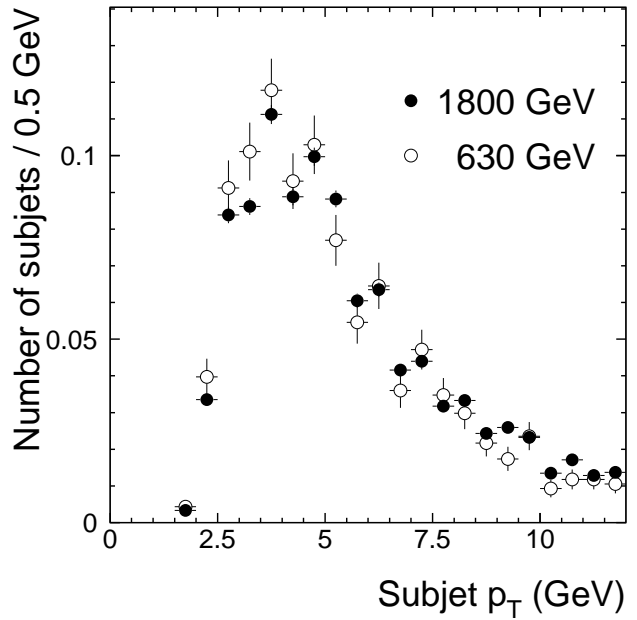}
\vspace{0.2cm}
\caption{
Same as in 
Fig.~\ref{fig:sub_et}, 
but with the low $p_T$ region expanded.
The increase at low $p_T$ is observed for all 
$y_{{\rm cut}}$,
but the specific cutoff at $p_T({\rm subjet}) \approx 1.75$ GeV 
is determined by our chosen value of $y_{{\rm cut}} = 10^{-3}$.
}
\label{fig:sub_loet}
\end{figure}

Subjets were defined through the 
product of their fractional jet $p_T$ 
and their separation in $(\eta,\phi)$ space 
[see Eqs.~(\ref{eq:dij}) and (\ref{eq:y})].
As shown in Figs.~\ref{fig:sub_et} and \ref{fig:sub_loet},
the shapes of the subjet $p_T$ spectra of the selected jets are
similar at the two beam energies.
The distributions suggest that jets are composed of a hard component 
and a soft component.
The peak at about 55 GeV and fall-off at higher $p_T$
is due to single-subjet jets and
the jet $p_T$ selections
($55 < p_T < 100$ GeV).
The threshold at subjet $p_T \approx 1.75$ GeV 
is set by the value $y_{{\rm cut}} = 10^{-3}$
and the minimum jet $p_T$ in the sample.

\begin{figure}
\epsfxsize=3.375in
\epsfbox{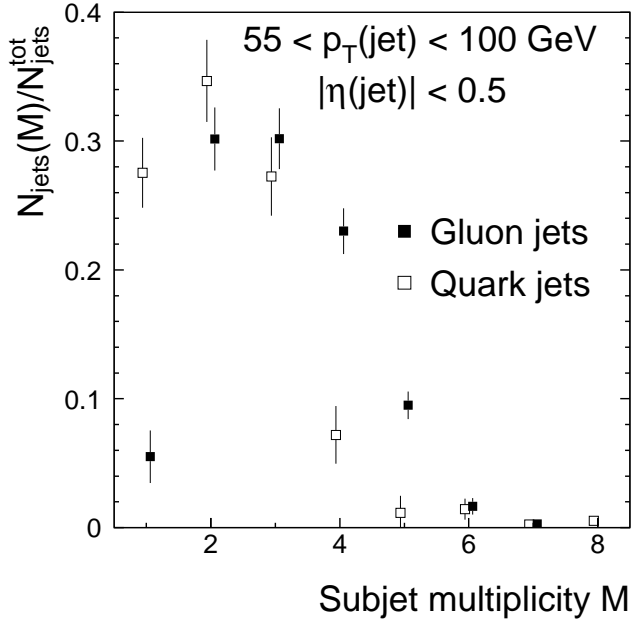}
\vspace{0.2cm}
\caption{
Uncorrected subjet multiplicity in gluon and quark jets, 
extracted from D\O\ data at
$\sqrt{s} = 1800$ GeV and 630 GeV,
using nominal gluon-jet fractions
$f_{1800} = 0.59$ and 
$f_{630} = 0.33$.
}
\label{fig:d0qg}
\end{figure}

While the $M_{1800}$ and $M_{630}$ inclusive measurements at $\sqrt{s} = 1800$ GeV and 
$\sqrt{s} = 630$ GeV are interesting in themselves,
they can be interpreted in terms of their gluon and quark content.
According to Eqs.~(\ref{eq:mg}) and (\ref{eq:mq})
the distributions in Fig.~\ref{fig:1800_630}
and their gluon-jet fractions at the two beam energies
can yield the uncorrected subjet multiplicity
distributions in gluon and quark jets.
The extracted measurements of $M_g$ and $M_q$
are shown in Fig.~\ref{fig:d0qg}
for the nominal values 
$f_{1800} = 0.59$ and
$f_{630} = 0.33$.  
As in the Monte Carlo simulation, the D\O\ data clearly indicate the
presence of
more subjets in gluon jets than in quark jets.
Such distributions can be used directly 
(without correcting the subjet multiplicities)
to discriminate between gluon and quark jets.
The results depend only on Monte Carlo estimates of
gluon-jet fractions at the two values of $\sqrt{s}$,
and not on any detailed simulation of jet structure.

\begin{figure}
\epsfxsize=3.375in
\epsfbox{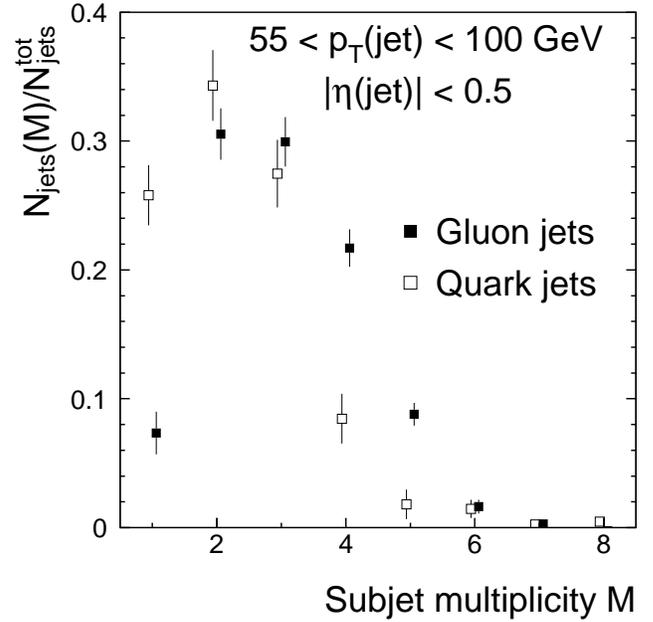}
\vspace{0.2cm}
\caption{
Uncorrected subjet multiplicity in gluon and quark jets, 
extracted from D\O\ data at
$\sqrt{s} = 1800$ GeV and 630 GeV,
using gluon-jet fractions
$f_{1800} = 0.61$ and 
$f_{630} = 0.30$.  
}
\label{fig:d0qg_61fg30}
\end{figure}

The sensitivity of $M_g$ and $M_q$
to the assumed values of $f_{1800}$ and
$f_{630}$ was checked by
investigating how the signal 
(i.e., the difference between $M_g$ and $M_q$)
depended on this choice.
It was found that when the 
gluon-jet fractions are either both increased or both decreased,
the signal remains relatively unchanged.
However, when
the gluon-jet fractions are changed
in opposite directions, this
produces the largest change in the
difference between gluon and quark jets.
The result of using
$f_{1800} = 0.61$ and $f_{630} = 0.30$,
instead of their nominal values,
is shown 
in the extracted distributions of Fig.~\ref{fig:d0qg_61fg30}.
The $M_g$ and $M_q$ distributions of Fig.~\ref{fig:d0qg_61fg30}
are qualitatively similar to those of Fig.~\ref{fig:d0qg},
and the large difference between gluon and quark jets 
is still apparent.

The subjet multiplicity distributions 
can be characterized by their means 
$\langle M \rangle$,
and by $\langle M \rangle - 1$, which correspond to
the average
number of subjet {\em emissions} in a gluon or quark jet.
For the nominal uncorrected D\O\ data
shown in Fig.~\ref{fig:d0qg},
$\langle M_g^{{\rm meas}} \rangle = 3.05 \pm 0.06$
and
$\langle M_q^{{\rm meas}} \rangle = 2.28 \pm 0.08$.
The analogous values for the Monte Carlo events
(see Fig.~\ref{fig:mc}) are
$\langle M_g^{{\rm meas}} \rangle = 3.01 \pm 0.09$
and
$\langle M_q^{{\rm meas}} \rangle = 2.28 \pm 0.08$.
Because the quoted statistical uncertainty 
on $\langle M_g^{{\rm meas}} \rangle$
is correlated with that on $\langle M_q^{{\rm meas}} \rangle$,
we define a ratio~\cite{aleph,Abreu:1998ve}
of emissions in gluon jets to quark jets:
\begin{equation}
r\equiv \frac{\langle M_{g} \rangle -1}
{\langle M_{q} \rangle -1}.
\label{eq:r}
\end{equation}
A value of $r = 1$ would mean that the substructure of gluon jets 
does not differ from that of quark jets.
The ratio has a value of $r = 1.61 \pm 0.15$ 
for the uncorrected data 
of Fig.~\ref{fig:d0qg},
and $r = 1.58 \pm 0.16$ for the analogous Monte Carlo events of 
Fig.~\ref{fig:mc},
where both uncertainties are statistical.
Using different
values for gluon-jet fraction at the two values of $\sqrt{s}$
(as in Fig.~\ref{fig:d0qg_61fg30}),
yields the range of $r$ values
given in Table~\ref{tab:fg}.
As expected, the observed ratio is smallest when the
fraction of gluon jets increases at 
$\sqrt{s} = 1800$ GeV and decreases at
$\sqrt{s} = 630$ GeV.
The two values of $f$ are the only assumptions from Monte Carlo, and
correspond to the largest source of systematic uncertainty
on $r$ (described more fully in Sec.~\ref{subs:sys}).
In all cases, we find that $r$ is significantly
greater than unity,
meaning that gluon jets and quark jets differ in their substructure.

\begin{table} \centering
\begin{tabular}{ccccc}
$f_{1800}$ & $f_{630}$ & $\langle M_g \rangle$ & $\langle M_q \rangle$ & $r$ \\
\hline
0.59 & 0.33 & 3.05 $\pm$ 0.06 & 2.28 $\pm$ 0.08 & 1.61 $\pm$ 0.15 \\ 
0.61 & 0.30 & 2.99 $\pm$ 0.05 & 2.34 $\pm$ 0.07 & 1.49 $\pm$ 0.11 \\ 
0.61 & 0.36 & 3.05 $\pm$ 0.06 & 2.24 $\pm$ 0.09 & 1.65 $\pm$ 0.16 \\ 
0.57 & 0.30 & 3.06 $\pm$ 0.06 & 2.31 $\pm$ 0.07 & 1.57 $\pm$ 0.14 \\ 
0.57 & 0.36 & 3.15 $\pm$ 0.08 & 2.19 $\pm$ 0.10 & 1.81 $\pm$ 0.22 \\
\end{tabular}
\caption{The uncorrected subjet multiplicity in gluon and quark jets,
and their ratio, extracted from D\O\ data,
assuming different values of gluon-jet fractions 
at the two center-of-mass energies,
based, in part, on Figs.~\ref{fig:isid34_cut_integ_100_big} and ~\ref{fig:gluon_frac_vs_etmin}.
}
\label{tab:fg}
\end{table}

\subsection{Corrected subjet multiplicity}
\label{subs:cor}
\begin{figure}
\epsfxsize=3.375in
\epsfbox{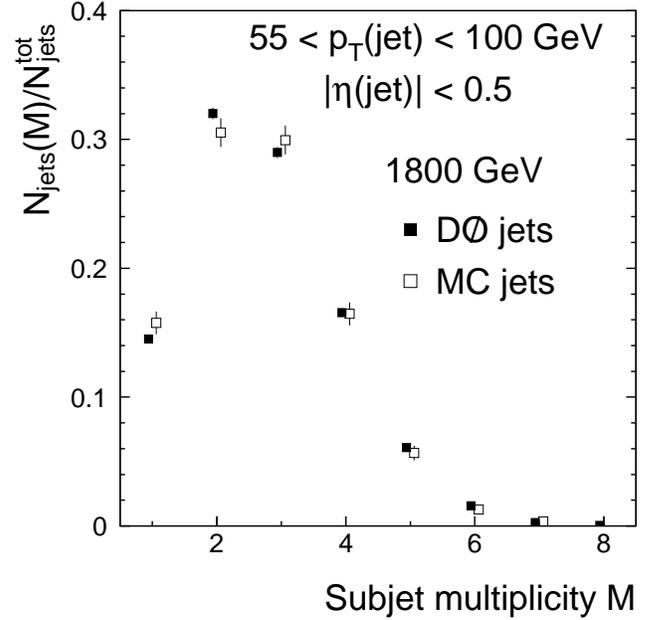}
\vspace{0.2cm}
\caption{
Uncorrected subjet multiplicity in jets from D\O\ 
and fully-simulated Monte Carlo events at 
$\sqrt{s} = 1800$ GeV.
}
\label{fig:a_d0mc}
\end{figure}

\begin{figure}
\epsfxsize=3.375in
\epsfbox{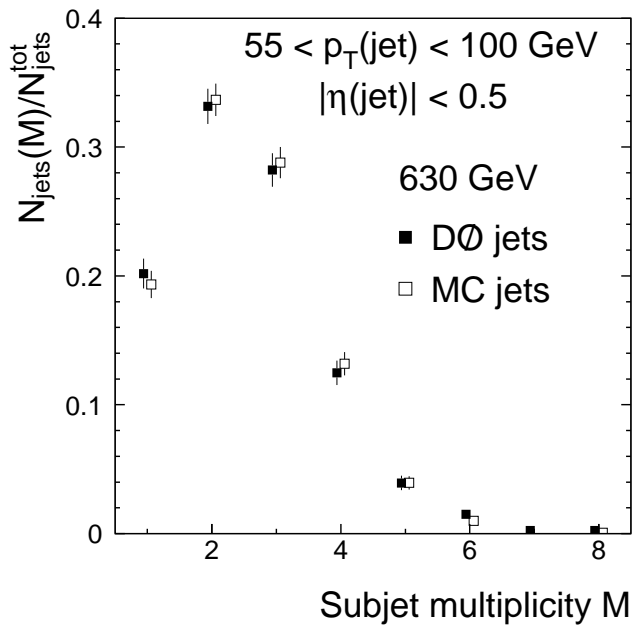}
\vspace{0.2cm}
\caption{Uncorrected subjet multiplicity in jets from D\O\ 
and from fully-simulated Monte Carlo events at 
$\sqrt{s} = 630$ GeV.
}
\label{fig:l_d0mc}
\end{figure}

As was stated above, the experimental conditions described in
Sec.~\ref{subs:jes} smear the measurement of the 
subjet multiplicity.
Although $r$ expresses differences between gluon and quark jets as a ratio
of mean subjet multiplicities,
the extracted $M_g$ and $M_q$ distributions
need separate corrections for the various detector-dependent effects that
can affect the value of $r$. 
The corrections are derived using Monte
Carlo events, which 
are in agreement with the uncorrected D\O\ data,
as shown in Figs.~\ref{fig:a_d0mc}
and \ref{fig:l_d0mc}.
The decomposition of the Monte Carlo events into $M_g$ and $M_q$
components was discussed in Sec.~\ref{subs:ext}.
The distributions shown in Fig.~\ref{fig:mc}
represent the uncorrected results for Monte Carlo events
that we use to derive the unsmearing corrections.

\begin{figure}
\epsfxsize=3.375in
\epsfbox{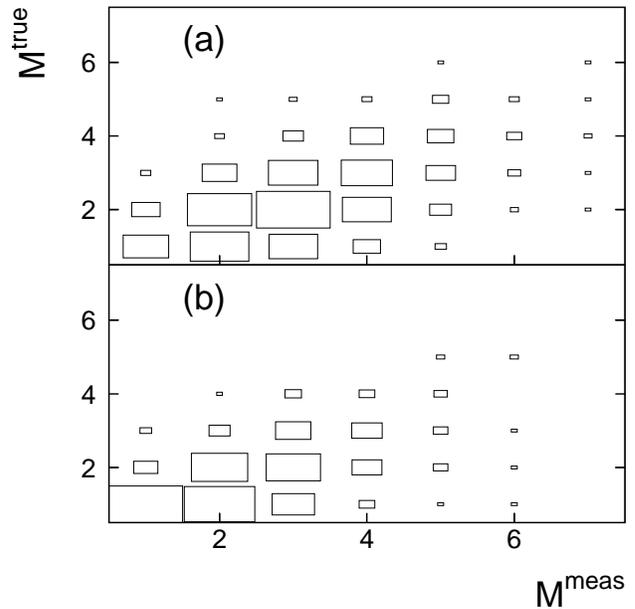}
\vspace{0.2cm}
\caption{The subjet multiplicity at particle-level ($M^{{\rm true}}$)
versus
the subjet multiplicity at
calorimeter-level ($M^{{\rm meas}}$) (includes effects of luminosity), 
at $\sqrt{s}=1800$ GeV,
for (a) gluon and (b) quark jets.}
\label{fig:uz_cal_vs_particle}
\end{figure}

The corrected distributions of $M_g$ and $M_q$
are defined in Monte Carlo jets at the particle level
(i.e., before development in the calorimeter).
All selected calorimeter-level jets are
matched (within $\Delta{\cal R} < 0.5$)
to jets reconstructed 
at the
particle level.
The matching procedure implicitly accounts for any
mismeasurement of jet $p_T$ because there is no
$p_T$ requirement in the matching.
The preclustering and clustering algorithms applied at
the particle level are identical to those applied
at the detector level. 
We tag simulated detector jets as either gluons or quarks,
and correlate the subjet multiplicity in particle jets ($M^{{\rm true}}$)
with that of detector partners ($M^{{\rm meas}}$).
These correlations are shown in Fig.~\ref{fig:uz_cal_vs_particle}
at $\sqrt{s} = 1800$ GeV,
and define the correction applied to the subjet multiplicity.
Similar results are available
at $\sqrt{s} = 630$ GeV (not shown).

The correction retrieves $M^{{\rm true}}$ from $M^{{\rm meas}}$,
in bins of $M^{{\rm meas}}$.
In general, the distributions of $M^{{\rm true}}_g$ and $M^{{\rm true}}_q$
in Fig.~\ref{fig:uz_cal_vs_particle}
are shifted to lower values 
relative to $M^{{\rm meas}}_g$ and $M^{{\rm meas}}_q$.
The shift in $M$ is due mainly to the effects of showering in the calorimeter,
rather than from
the combined effects of multiple interactions, pile-up, and uranium noise,
which are reduced by using $D = 0.5$.
Fortunately, shower development is independent 
of beam energy,
and the other contributions differ only slightly
(see Sec.~\ref{subs:sys}).

Shower development in the 
calorimeter tends to add subjets to a jet
because any single particle can deposit energy
in several towers of the calorimeter.
Signals in many towers 
generate a large number of preclusters,
and in turn, a large number of subjets.
However, the opposite can also occur.
For example, 
when two subjets at the particle level 
(each composed of one or two hadrons)
deposit energy in a region of the calorimeter between them,
such energy can ``bridge'' 
distinct subjets at the particle level into a single
subjet at the calorimeter level.  This bridging effect 
is more pronounced in jets that already have a large $M^{{\rm true}}$.
For this reason, the effects of multiple interactions, 
pile-up, and uranium noise
tend to reduce the correction to $M^{{\rm meas}}$.

\begin{figure}
\epsfxsize=3.375in
\epsfbox{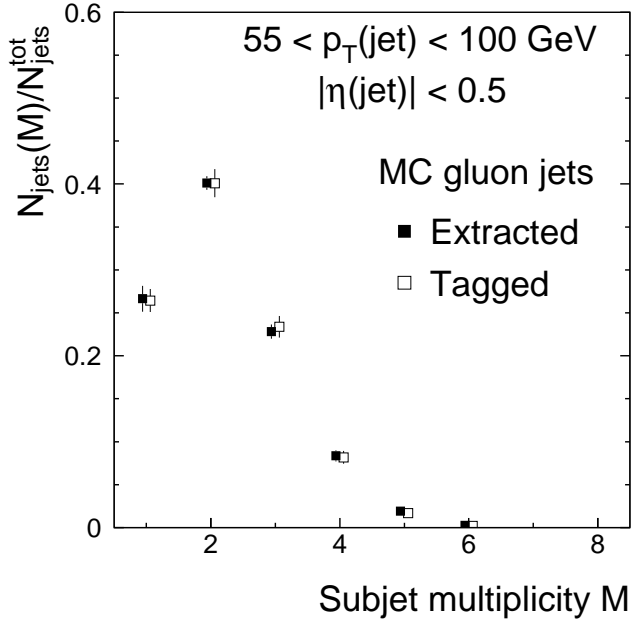}
\vspace{0.2cm}
\caption{
The subjet multiplicity in Monte Carlo gluon jets.
The extracted distribution has been unsmeared.
The tagged distribution was obtained directly from particle-level
gluon jets at  $\sqrt {s}=1800$ GeV.
}
\label{fig:g_calpcl}
\end{figure}

\begin{figure}
\epsfxsize=3.375in
\epsfbox{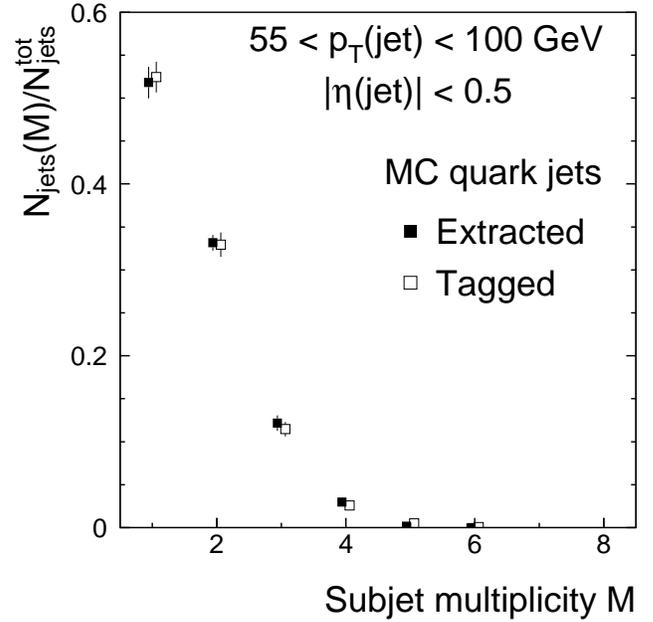}
\vspace{0.2cm}
\caption{
The subjet multiplicity in Monte Carlo quark jets.
The extracted distribution has been unsmeared.
The tagged distribution was obtained directly from particle-level
quark jets at  $\sqrt {s}=1800$ GeV.
}
\label{fig:q_calpcl}
\end{figure}

To check that the correction defined by 
the correlations in
Fig.~\ref{fig:uz_cal_vs_particle} is valid,
it was applied to the uncorrected
$M_g$ and $M_q$ Monte Carlo distributions 
in Fig.~\ref{fig:mc}.
The resulting corrected distributions for
$M_g$ and $M_q$
are given in Fig.~\ref{fig:g_calpcl}
and Fig.~\ref{fig:q_calpcl}, respectively. 
The correction reduces the average subjet multiplicity
in the Monte Carlo to
$\langle M_{g}^{{\rm true}} \rangle = 2.19 \pm 0.04$
and
$\langle M_{q}^{{\rm true}} \rangle = 1.66 \pm 0.04$
and the corrected ratio is
$r = 1.82 \pm 0.16$.
Any remaining small differences between the extracted and the tagged
$M^{{\rm true}}$ distributions in
Fig.~\ref{fig:g_calpcl} and Fig.~\ref{fig:q_calpcl}
are attributable to the differences between the extracted
and the tagged $M^{{\rm meas}}$ (at $\sqrt{s} = 1800$ GeV) of 
Fig.~\ref{fig:mc}.
These differences are smaller for the corrected
distributions $M^{{\rm true}}$, 
than for
the uncorrected distributions.

\begin{figure}
\epsfxsize=3.375in
\epsfbox{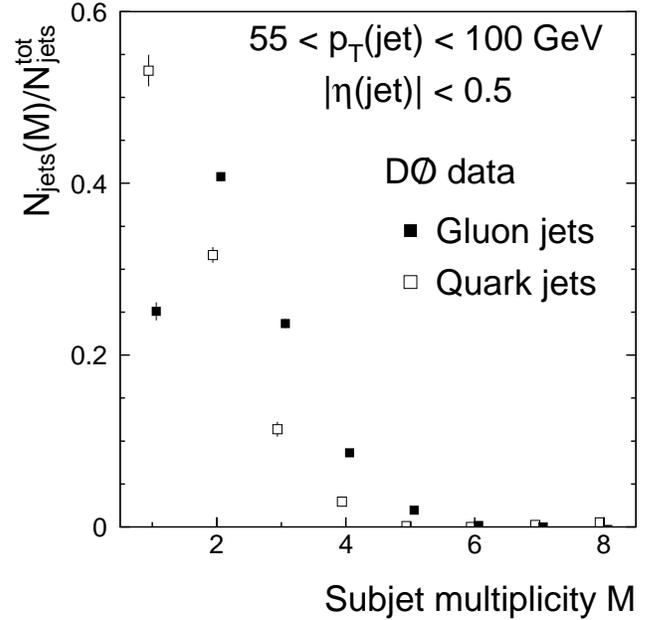}
\vspace{0.2cm}
\caption{Corrected subjet multiplicity for gluon and quark jets, 
extracted from D\O\ data.
}
\label{fig:qg}
\end{figure}

Figure~\ref{fig:qg} shows the corrected subjet multiplicities
for gluon and quark jets.
The rate for $M = 1$ quark jets has almost doubled,
while the rate for $M = 3$ quark jets has fallen by a factor
of $\approx 2$,
relative to the uncorrected result.
A similar effect is observed for gluon jets.
From Fig.~\ref{fig:qg}, we obtain the corrected
mean values in the D\O\ data to be
$\langle M_{g}^{{\rm true}} \rangle = 2.21 \pm 0.03$
and
$\langle M_{q}^{{\rm true}} \rangle = 1.69 \pm 0.04$,
which gives $r = 1.75 \pm 0.15$,
in good agreement with the prediction from {\sc HERWIG}.
The unsmearing therefore widens the difference between 
gluon and quark jets.

We choose not to correct $M$ for any impact of
the preclustering algorithm on subjet multiplicity.
Instead, the preclustering algorithm can be applied easily
to the particle-level events in Monte Carlo,
and these are therefore treated in the same way as the D\O\ data. 
For completeness, we note that $r$ can decrease by as much as
0.2 at the particle level, when preclustering is turned off.

\subsection{Additional corrections and systematic uncertainties}
\label{subs:sys}
The dominant systematic uncertainty on the subjet multiplicity
arises from the uncertainty on the gluon-jet fractions.
In fixed-order perturbative QCD,
the jet cross section at any given $p_T$ 
is a more-steeply-falling function of $p_T$
at $\sqrt{s} = 630$ GeV than at $\sqrt{s} = 1800$ GeV~\cite{iain}.
Consequently, applying identical cutoffs
biases the
$\langle p_T \rangle$ of jets  
at $\sqrt{s} = 1800$ GeV upwards relative 
to $\sqrt{s} = 630$ GeV.
Monte Carlo studies indicate this bias is approximately $2$ GeV.
One way to compensate for this effect is to
shift the $p_T$ range at $\sqrt{s} = 630$ upwards by a few GeV.
Due to the steep negative slope of the jet-$p_T$ spectrum,
it is sufficient to shift only the lower edge of the $p_T$ bins.
When this is done, Fig.~\ref{fig:isid34_cut_integ_100_big} shows
that the change in gluon-jet fraction is $\Delta f  < 0.03$.
We do not correct $f$ for this, but 
account for this residual effect in the
systematic uncertainty associated with the 
jet $p_T$.

Changing the gluon-jet fractions used in the analysis
gives a direct estimate of the uncertainty on the subjet multiplicity.
We will motivate the range of uncertainty in gluon-jet fractions
at the two center-of-mass energies by investigating the behavior of the
PDFs. For the jet samples used in this analysis,
the average jet $p_T$ was approximately 
65 GeV.
This jet $p_T$ probes an average
$x$ value of 0.07 at $\sqrt{s}=1800$ GeV
and 0.2 at $\sqrt{s}=630$ GeV.
In these regions of $x$, the quark PDFs are well-constrained
by existing data.
However, the gluon PDF is not so well-constrained.
We examined different
parameterizations of the gluon PDF
at the two
$x$ values of interest.
In particular, the MRST5~\cite{mrs} gluon PDF is $21\%$ smaller than
the CTEQ4M parameterization at $x = 0.2$, but only
$4\%$ smaller at $x = 0.07$.
This and other comparisons between PDFs show
larger fractional differences at $x = 0.2$ than at
$x = 0.07$.

Assuming that the quark distributions are essentially identical
in different PDF 
parameterizations, the gluon-jet fraction $f$
for different PDFs can be estimated as
\begin{equation}
f = \frac {f^{{\rm ref}} + \epsilon f^{{\rm ref}}}{(f^{{\rm ref}} + \epsilon f^{{\rm ref}}) + (1 - f^{{\rm ref}})}
\label{eq:fprime}
\end{equation}
where $f^{{\rm ref}}$ is the gluon-jet fraction from some reference PDF,
and $\epsilon$ is a fractional difference in the gluon PDF.
Table~\ref{tab:f_pdf} shows the gluon-jet fractions
estimated for PDFs at the two center-of-mass energies.
The MRST5 set shows the largest departure relative to CTEQ4M.
In all cases, the change in $f$ 
is in the same direction at both $\sqrt{s}$.

\begin{table} \centering
\begin{tabular}{lrlcr@{.}lc}
PDF set &\multicolumn{1}{c}{$\sqrt{s}$ (GeV)}&\multicolumn{1}{c}{$x$}& $x g(x)$ &\multicolumn{2}{c}{$\epsilon$}& $f_{\sqrt{s}}$ \\
\hline
CTEQ4M  & 1800 & 0.07 & 1.643 &   0  & 00 & 0.59 \\ 
CTEQ4HJ & 1800 & 0.07 & 1.643 &   0  & 00 & 0.59 \\ 
CTEQ2M  & 1800 & 0.07 & 1.714 &   0  & 04 & 0.60 \\ 
CTEQ5M  & 1800 & 0.07 & 1.614 & $-0$ & 02 & 0.59 \\ 
CTEQ5HJ & 1800 & 0.07 & 1.586 & $-0$ & 04 & 0.58 \\ 
MRST5   & 1800 & 0.07 & 1.586 & $-0$ & 04 & 0.58 \\ 
GRV94   & 1800 & 0.07 & 1.743 &   0  & 06 & 0.60 \\ 
\hline
CTEQ4M  &  630 & 0.2  & 0.365 &   0  & 00 & 0.33 \\ 
CTEQ4HJ &  630 & 0.2  & 0.340 & $-0$ & 06 & 0.32 \\ 
CTEQ2M  &  630 & 0.2  & 0.385 &   0  & 06 & 0.34 \\ 
CTEQ5M  &  630 & 0.2  & 0.340 & $-0$ & 06 & 0.32  \\ 
CTEQ5HJ &  630 & 0.2  & 0.350 & $-0$ & 03 & 0.32 \\ 
MRST5   &  630 & 0.2  & 0.290 & $-0$ & 21 & 0.28 \\ 
GRV94   &  630 & 0.2  & 0.405 &   0  & 12 & 0.36 \\ 
\end{tabular}
\caption{
Values of gluon-jet fractions
for different PDFs, calculated using 
Eq.~(\ref{eq:fprime}), at a jet $p_T = 65$ GeV.
The CTEQ4M parameterization is chosen as the reference.
The fractional change in the gluon PDF $g(x)$ is given by
$\epsilon = (g(x) - g^{{\rm ref}}(x))/g^{{\rm ref}}(x)$, where $g^{{\rm ref}}(x)$
is the reference.
}
\label{tab:f_pdf}
\end{table}

The preceeding discussion assumed that 
the PDFs had
the same quark distribution.
In reality, the quark PDFs also tend to change when the gluon PDF changes.
When this compensating effect is taken into account in Eq.~(\ref{eq:fprime}),
the equivalent MRST5 gluon-jet fractions become $f_{1800} = 0.58$
and $f_{630} = 0.29$.

Based on the above, we assign uncertainties to the gluon-jet fractions
of $\pm 0.02$ at $\sqrt{s}=1800$, and $\pm 0.03$ at $\sqrt{s}=630$.  
In fact, we vary the gluon-jet fraction
in opposite directions, using
$f_{1800} = 0.61$ and $f_{630} = 0.30$, and
$f_{1800} = 0.57$ and $f_{630} = 0.36$,
to gauge the impact on $r$.
As in Sec.~\ref{subs:raw}, we repeat the analysis 
assuming these different input gluon-jet
fractions, this time including the
correction to the particle level. 
The extracted ratios are summarized
in Table~\ref{tab:sys_fg}. 
The largest departures from the reference value of $r=1.75$
define the systematic uncertainties of
$\pm^{0.17}_{0.10}$.

The second-largest source of systematic uncertainty in the subjet multiplicity
stems from an uncertainty in the measurement
of jet $p_{T}$. 
A mismeasurement of jet $p_{T}$ will lead to the selection of a
slightly different sample of jets, but will not affect the subjet
multiplicity directly. 
If jet $p_{T}$ is mismeasured at both center-of-mass 
energies, we expect the effect to partially cancel in the
ratio $r$. 
An estimate of the impact from this uncertainty
is therefore obtained by varying the jet $p_{T}$ only
at $\sqrt{s}=1800$ GeV. 
Since the calorimeter response is
independent of $\sqrt{s}$,
we
estimate the effect of a difference in any offset in $p_{T}$ at the
two center-of-mass energies by
changing the jet-$p_{T}$ window from 
$55<p_{T}<100$ GeV to  $57<p_{T}<100$ GeV at $\sqrt{s}=1800$ GeV. 
A 2 GeV shift in the measured jet $p_T$ corresponds approximately to
two times the total
offset $p_O$ for $k_{\perp}$ jets reconstructed with
$D = 0.5$.  [This assumes $p_O(D)$ scales as $D^2 p_O(D=1.0)$].
This reduces the subjet multiplicity ratio $r$ by 0.12,
which is taken as a
symmetric systematic uncertainty.

\begin{table} [t] 
\centering
\begin{tabular}{ccccc}
$f_{1800}$ & $f_{630}$ & $\langle M_g \rangle$ & $\langle M_q \rangle$ & $r$ \\
\hline
0.59 & 0.33 & 2.21 $\pm$ 0.03 & 1.69 $\pm$ 0.04 & 1.75 $\pm$ 0.15 \\ 
0.61 & 0.30 & 2.18 $\pm$ 0.02 & 1.72 $\pm$ 0.04 & 1.65 $\pm$ 0.12 \\ 
0.61 & 0.36 & 2.20 $\pm$ 0.03 & 1.67 $\pm$ 0.05 & 1.79 $\pm$ 0.17 \\ 
0.57 & 0.30 & 2.21 $\pm$ 0.03 & 1.70 $\pm$ 0.04 & 1.72 $\pm$ 0.14 \\ 
0.57 & 0.36 & 2.24 $\pm$ 0.04 & 1.65 $\pm$ 0.05 & 1.92 $\pm$ 0.22 \\ 
\end{tabular}
\caption{Subjet multiplicity in gluon and quark jets,
and their ratio,
extracted from D\O\ data and corrected to the particle level,
assuming different gluon-jet fractions
at the two center-of-mass energies.
}
\label{tab:sys_fg}
\end{table}

Because the correction to the particle level
produces a large change in the
shape of the subjet multiplicity distribution, 
we will estimate the impact of the
unsmearing on the systematic uncertainty on $r$.
This uncertainty has two parts:
one is the uncertainty due to the simulation of effects arising 
from dependence on luminosity,
and the other is the uncertainty in the simulation of the D\O\ calorimeter.
To account for the former, we use an alternate 
Monte Carlo sample at $\sqrt{s} = 630$ GeV,
with a luminosity of 
${\cal L} \approx 0.1 \times 10^{30} \text{cm}^{-2}\text{s}^{-1}$, 
and note that $r$ increases by 0.13.
Such a small change in $r$ indicates that it
depends only weakly
on luminosity.  Nevertheless, we increase
our nominal value of
$r = 1.75$ by half of the difference (to $r = 1.82$),
and take this correction as a symmetric
systematic uncertainty of $\pm$ 0.07.

To evaluate the other part of the uncertainty on the unsmearing,
we compare two types of simulations of the D\O\ calorimeter.
The default fast simulation ({\small SHOWERLIB})
is a library that contains single-particle calorimeter showers
obtained using the {\small GEANT} full detector simulation.
{\small SHOWERLIB}
truncates the number of calorimeter cells associated with
each individual particle, but rescales the energy
of the shower to agree with the average energy given by the full {\small GEANT}
simulation.
The full {\small GEANT} simulation, while slower, 
accounts for the precise geometry of the uranium plates in the calorimeter
and has no truncation.
In a test using a limited number of Monte Carlo events, the latter simulation
produced more subjets than the former,
and so we increase the value of the ratio
by 0.02 (half the difference of the $r$ values in 
each simulation)
to $r=1.84$, and take this correction as another symmetric
systematic error of $\pm$ 0.02. 
Applying the same additional corrections to the nominal ratio in the 
Monte Carlo 
gives a final result of $r = 1.91$ for {\sc HERWIG}.

\begin{table}
\setlength{\tabcolsep}{1.5pc}
\newlength{\digitwidth} \settowidth{\digitwidth}{\rm 0}
\catcode`?=\active \def?{\kern\digitwidth}
\begin{tabular}{@{}lr}
Source		 	& $\delta r$ \\
\hline
Gluon-jet fraction    	& $^{+0.17}_{-0.10}$ \\
Cutoff on jet $p_T$	& $\pm 0.12$ \\
Unsmearing          	& $\pm 0.07$ \\
Detector simulation   	& $\pm 0.02$ \\
\hline 
Total         	        & $^{+0.22}_{-0.18}$ \\
\end{tabular}
\caption{Systematic uncertainties on the ratio $r$.}
\label{tab:sys}
\end{table}

A list of the systematic uncertainties is shown in Table~\ref{tab:sys}, 
all of which
are added in quadrature to 
obtain the total uncertainty of the corrected ratio. 
The final result for the ratio is 
\begin{equation}
r\equiv \frac{\langle M_{g} \rangle -1} {\langle M_{q} \rangle -1}
=1.84\pm
0.15\;(\text{stat.})\pm ^{0.22}_{0.18} \;(\text{sys.}).
\end{equation}

\section{Conclusion}

\label{subs:conc}

We present two analyses 
of D\O\ data
using
the $k_{\perp}$ jet reconstruction algorithm.
One 
analysis examines the $p_T$
and direction of $k_{\perp}$ jets
reconstructed with the parameter $D = 1.0$.
For this measurement of the jet $p_T$ spectrum,
we describe 
a procedure to calibrate the
momentum of $k_{\perp}$ jets
based on our experience with the cone algorithm,
but 
using 
an improved technique
for determining
the offset correction.
Compared to our published results
for the cone algorithm with ${\cal R} = 0.7$~\cite{jes},
the $k_{\perp}$ jet algorithm
with $D = 1.0$ reconstructs $40-50\%$ more energy from uranium noise,
pile-up, multiple $p\bar{p}$ interactions, and the underlying event,
and has a smaller uncertainty on the offset.
We also report the results of a direct comparison of
the $k_{\perp}$ and cone 
algorithms, on an event-by-event basis.
Considering only the two leading jets in the central region ($|\eta| < 0.5$),
the $k_{\perp}$ and cone 
jet axes coincide within $\Delta{\cal R} =$ 0.1 (0.5)
at the 91$\%$ (99.94$\%$) level.
Matching with $\Delta{\cal R} =$ 0.5,
the corrected $p_T$ of $k_{\perp}$ jets is higher than the corrected
$E_T$ of cone jets.
The difference is roughly linear in jet $p_T$, 
varying from about $5$ GeV at $p_T \approx 90$ GeV 
to about $8$ GeV at $p_T \approx 240$ GeV.

\begin{figure}
\epsfxsize=3.375in
\epsfbox{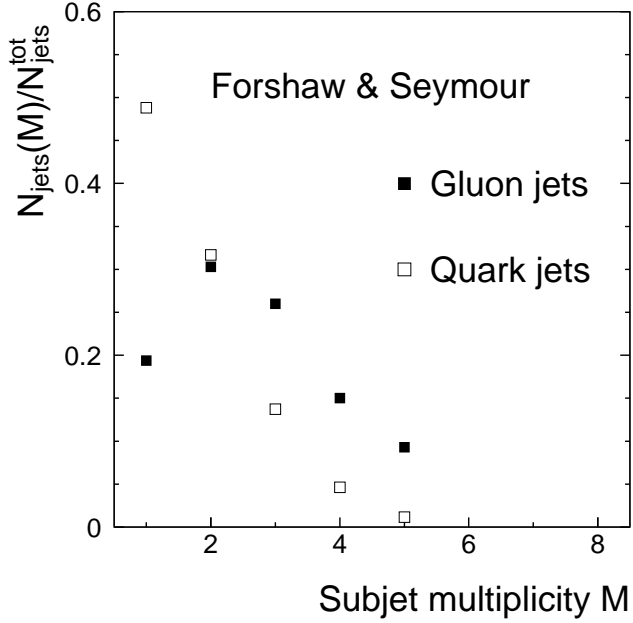}
\vspace{0.2cm}
\caption{
The subjet multiplicity in gluon and quark jets,
for $y_{{\rm cut}} = 10 ^{-3}$
(as defined by Eq.~[\ref{eq:y}]),
in a resummation calculation by Forshaw and 
Seymour \protect\cite{sey_sub2}.
The jets are produced at $\sqrt{s} = 1800$ GeV,
with $p_T = 65$ GeV and $\eta = 0$,
using the CTEQ4M PDF,
and are reconstructed with $D = 0.5$.
The points in the fifth bin 
refer to $M \geq 5$.
}
\label{fig:sey_sub2_page18}
\end{figure}

\begin{figure}
\epsfxsize=3.375in
\epsfbox{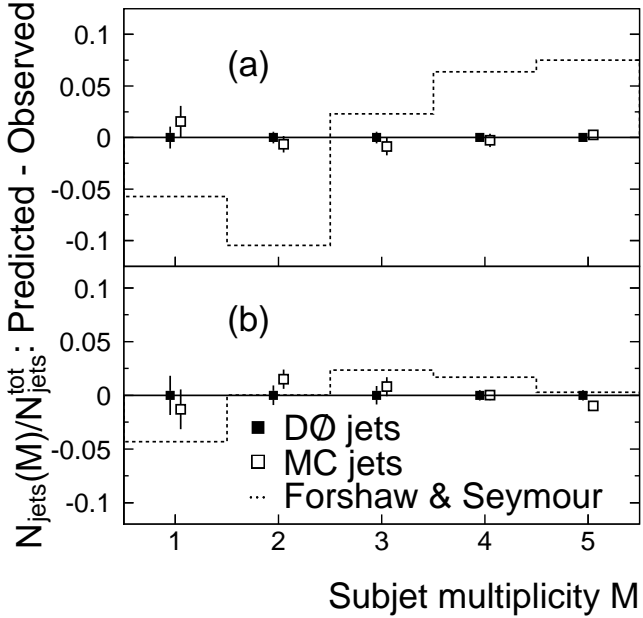}
\vspace{0.2cm}
\caption{
The subjet multiplicity in (a) gluon and (b) quark jets,
for D\O\ data, for the {\sc HERWIG} Monte Carlo, and 
resummed predictions.
The resummed prediction does not use a preclustering algorithm. 
The points in the fifth bin 
are for $M \geq 5$.
The D\O\ data (see Fig.~\ref{fig:qg})
have been subtracted
from each set of points.
}
\label{fig:d0seyhw_trunc}
\end{figure}

In the other analysis,
we probe the structure of
central $k_{\perp}$ jets 
reconstructed with the parameter $D = 0.5$,
and find that the
{\sc HERWIG} 
Monte Carlo predictions of subjet multiplicity 
are in excellent agreement with our measurements.
The subjet multiplicities in gluon and quark jets, 
predicted by a
fully resummed calculation~\cite{sey_sub2},
and shown in Fig.~\ref{fig:sey_sub2_page18},
are qualitatively consistent with our data,
but their mean values are slightly high.
This discrepancy may be due to
the fact that the calculation lacks a preclustering algorithm.  
The subjet multiplicity distributions,
where we have subtracted
the D\O\ values from the predictions,
are shown in Fig.~\ref{fig:d0seyhw_trunc}.
The ratio of mean multiplicities for
the resummed calculation 
(which assumes $M \leq 5$)
is $r = 2.12$.  
The ratio in the D\O\ data increases by 0.06 with the
assumption $M \leq 5$. Therefore, the resummed prediction
is well within the limits of experimental uncertainty.
The ratio measured at D\O\ agrees with the result
of $r = 1.7 \pm 0.1$ from ALEPH, measured in $e^+e^-$
annihilations at $\sqrt{s} = M_Z$ for a subjet
resolution parameter $y_{o} = 10^{-3}$~\cite{aleph},
and with the associated Monte Carlo and 
resummation prediction~\cite{sey_sub0},
but is higher than the ratio measured at DELPHI~\cite{Abreu:1998ve}.
The DELPHI result uses a different definition of the
jet resolution scale $y$ than used by ALEPH ($y_1$), 
which takes the place of $D$ in a hadron collider,
making direct comparisons difficult.
These experimental and theoretical values for
$r$ 
are all smaller than the naive QCD prediction of the ratio 
of color charges of 2.25.
This may be caused by higher-order radiation in QCD,
which tends to reduce the ratio from the naive value.

In summary, we present the first detailed
measurements of properties of $k_{\perp}$ jets
in hadron collisions.
Using the standard value $D = 1.0$ of the jet-separation
parameter in the $k_{\perp}$ algorithm,
we find that the
$p_T$ of $k_{\perp}$ jets is higher than the
$E_T$ of matched cone jets (with ${\cal R} = 0.7$)
by about $5$ (8) GeV at $p_T \approx 90$ (240) GeV.
To analyze internal jet structure,
we measure the multiplicity distribution
of subjets in $k_{\perp}$ jets with $D=0.5$
at $\sqrt{s} = 1800$ GeV and 630 GeV.
Exploiting the difference in gluon-jet fractions
at the two center-of-mass energies,
we extract the subjet multiplicity in gluon
and quark jets.
The differences between gluon and quark
jets are summarized in the ratio of average
emitted subjet multiplicities, measured as:
\begin{equation}
r\equiv \frac{\langle M_{g} \rangle -1} {\langle M_{q} \rangle -1}
=1.84 \pm ^{0.27}_{0.23}.
\end{equation}
The D\O\ result demonstrates that 
gluon and quark jets are significantly different
in hadron collisions,
and that it may be possible
to discriminate between them on an 
individual basis.

\section{Acknowledgments}

We thank Mike Seymour and Jeff Forshaw for
many useful discussions and
their assistance with the theoretical calculations.
%
We thank the staffs at Fermilab and collaborating institutions, 
and acknowledge support from the 
Department of Energy and National Science Foundation (USA),  
Commissariat  \` a L'Energie Atomique and 
CNRS/Institut National de Physique Nucl\'eaire et 
de Physique des Particules (France), 
Ministry for Science and Technology and Ministry for Atomic 
   Energy (Russia),
CAPES and CNPq (Brazil),
Departments of Atomic Energy and Science and Education (India),
Colciencias (Colombia),
CONACyT (Mexico),
Ministry of Education and KOSEF (Korea),
CONICET and UBACyT (Argentina),
The Foundation for Fundamental Research on Matter (The Netherlands),
PPARC (United Kingdom),
Ministry of Education (Czech Republic),
and the A.P.~Sloan Foundation.
%


\end{document}